\documentclass{aastex}
\usepackage{emulateapj5}

\newcommand{\etal}{{\it et\thinspace al.}\ }

\submitted{Submitted 22 September 2003; Accepted 2 January 2004 -- To appear in the April, 2004 AJ}

\lefthead{Van Duyne \etal }
\righthead{Low-Frequency KISS}

\begin{document}

\title{Radio Continuum Emission at 1.4 GHz from KISS Emission-Line Galaxies}

\author{Jeffrey Van Duyne\altaffilmark{1}, Eli Beckerman\altaffilmark{2}, John J. Salzer}
\affil{Astronomy Department, Wesleyan University, Middletown, CT 06459; slaz@astro.wesleyan.edu}

\author{Caryl Gronwall}
\affil{Department of  Astronomy \& Astrophysics, Pennsylvania State University, University Park, PA 16802; caryl@astro.psu.edu}

\author{Trinh X. Thuan}
\affil{Astronomy Department, University of Virginia, Charlottesville, VA 22903; txt@starburst.astro.virginia.edu}

\author{J. J. Condon}
\affil{National Radio Astronomy Observatory, Charlottesville, VA 22903; jcondon@nrao.edu}

\and
\author{Lisa M. Frattare}
\affil{Space Telescope Science Institute, Baltimore, MD 21218; frattare@stsci.edu}

\altaffiltext{1}{present address: Department of Astronomy, Yale University, New Haven, CT 06520;
vanduyne@astro.yale.edu }

\altaffiltext{2}{present address: Harvard Smithsonian Center for Astrophysics, 60 Garden Street, Cambridge, MA 02138;
eli@head-cfa.harvard.edu }

\begin{abstract}

We have searched the Faint Images of the Radio Sky at Twenty centimeters (FIRST) 
and the NRAO VLA Sky Survey (NVSS) 1.4 GHz radio surveys for sources that are
coincident with emission-line galaxy (ELG) candidates from the KPNO International 
Spectroscopic Survey (KISS).  A total of 207 of the 2157 KISS ELGs ($\sim$10\%) in the 
first two H$\alpha$-selected survey lists were found to possess radio detections in FIRST 
and/or NVSS.  Follow-up spectra exist for all of the radio detections, allowing us
to determine the activity type (star-forming vs. AGN) for the entire sample.
We explore the properties of the radio-detected KISS galaxies in order to gain a
better insight into the nature of radio-emitting galaxies in the local universe (z $<$ 0.1).
No dwarf galaxies were detected, despite the large numbers of low-luminosity galaxies
present in KISS, suggesting that lower mass, lower luminosity objects do not possess
strong galaxian-scale magnetic fields.  Due to the selection technique used for KISS,
our radio ELGs represent a quasi-volume-limited sample, which allows us to develop a clearer 
picture of the radio galaxy population at low redshift.  Nearly 2/3rds of the KISS radio
galaxies are starburst/star-forming galaxies, which is in stark contrast to the results
of flux-limited radio surveys that are dominated by AGNs and elliptical galaxies (i.e., 
classic radio galaxies).  While there are many AGNs among the KISS radio galaxies, 
there are no objects with large radio powers in our local volume.  We 
derive a radio luminosity function (RLF) for the KISS ELGs that agrees very well with 
previous RLFs that adequately sample the lower-luminosity radio population.

\end{abstract}

\keywords{galaxies: starburst --- galaxies: active ---
  galaxies:luminosity function --- radio continuum}

\section{Introduction}

Radio source surveys provide an unobstructed window for studying the
distant universe.   In the past few decades there have been many
attempts to construct radio luminosity functions for both radio- and
optically-selected surveys of galaxies \citep{c84,wind85,kron85,c89,ccb02}.  
Recently, studies have focused on more specific aspects of galaxian radio
emission by either trying to obtain samples of radio sources that reach 
to sub-mJy flux levels \citep{benn93,grupp99,george99,richards00,pran01}, 
or attempting to select radio galaxies from specialized subsamples 
\citep{brink00,ho01,yrc01}.  However, the focus of most of these studies 
has been on higher redshifts.  This comes about naturally due to the use 
of flux-limited samples of objects.  These samples are relevant for 
studying the evolution of the radio luminosity function.  In the current 
study we take a different approach, seeking to investigate the radio 
emission from a quasi-volume-limited sample of galaxies.  We explore the 
population of radio-emitting galaxies in the local universe
($z<0.1$) by selecting these galaxies from the KPNO International
Spectroscopic Survey (KISS; \cite{s00,s01}), which detects 
galaxies via the H$\alpha$ emission line.

Previous objective-prism galaxy surveys, such as the Markarian 
\citep{markarian67,mark81}, Tololo \citep{smith75,smith76}, Michigan 
\citep{michigan77,michigan81}, and Case \citep{cwru83,cwru92} surveys, 
have been major sources of objects from which we have learned much of
what we know about Seyfert galaxies, starburst galaxies and even QSOs.
These active galactic nuclei (AGNs) and galaxian starbursts are among
the most energetic phenomena known.  Quasars, by definition, are associated 
with intense radio emission.  However, the fraction of the other types of 
active galaxies that have associated radio emission 
remains uncertain.  Also in question is the fraction of less energetic or 
``normal'' galaxies that have detectable radio emission.  These galaxies 
may be the population that the very deep radio flux (i.e., sub-mJy level)
surveys detect.  Some observations in this regime \citep{grupp99,pran01} 
suggest that the faint radio galaxy population consists of far more ``normal,'' 
low radio power galaxies, which are more often starburst or star-forming 
galaxies rather than AGNs.  This is in direct contrast to mJy-level surveys 
that are dominated by radio-loud galaxies.  If this is so, these ``normal'' 
radio galaxies most likely exist in the local universe, and would make up a 
substantial fraction of a volume-limited radio sample.  

Recent studies \citep{pran01,yrc01,sad02} have used large-area optical
and infrared surveys (ESO Imaging Survey, IRAS, 2dF Galaxy Redshift survey) 
to identify galaxies with radio sources and construct radio-galaxy samples.  
We utilize KISS as the selection mechanism for a
local sample of radio emission galaxies which are positionally matched 
with radio sources identified by the combined catalogs of the Faint Images
of the Radio Sky at Twenty-centimeters (FIRST) survey \citep{beck95} and 
the NRAO VLA Sky Survey (NVSS) \citep{c98} at 1.4 GHz.  The objective-prism 
survey method of KISS automatically assigns redshift estimates to the 
identified radio galaxies, and follow-up spectroscopy of these galaxies
supplies the survey with activity types.  Lack of follow-up spectroscopy and
redshifts is the single largest obstacle to interpreting the contents of
most deep radio surveys.  Highly successful studies, such as the Marano Field
study conducted by \citet{grupp99}, were conducted with only 45\% of the
sub-mJy sample of galaxies having spectroscopic information.  More recent 
studies like \citet{pran01} have achieved almost 100\% follow-up spectroscopy 
for their galaxies.

The combination of the good depth and very specific wavelength coverage of the 
KISS objective-prism spectra, combined with the moderately deep brightness limits 
of FIRST ($S_{lim} = 1$ mJy/beam) and NVSS ($S_{lim} \approx 2.5$ mJy/beam) gives 
us a unique sample of local radio-emitting galaxies that has a high degree of 
completeness.   Furthermore, we sample intermediate and lower radio powers better 
than most previous studies.  This is an alternative method to probing into the nature 
of the so-called sub-mJy population which other groups \citep{george99,pran01} believe 
is dominated by star-forming galaxies (which KISS is sensitive to), as opposed to the
radio-loud objects found typically in galaxies with early-type morphologies.

This paper is organized as follows.  In \S2 we briefly summarize the surveys
used in the creation of this sample (KISS, FIRST, NVSS), while in \S3 we
discuss the optical-radio identification method, present a comparison 
between the FIRST and NVSS flux densities, and define the complete radio-KISS 
sample.  The radio-KISS sample properties are described in \S4
as well as the properties of KISS ELGs \emph{not} identified with radio
sources.  Section 5 presents the local 1.4 GHz radio luminosity function
derived from the radio-KISS sample and \S6 presents our conclusions.

\section{The Galaxy Sample}

\subsection{KPNO International Spectroscopic Survey - KISS}
\label{KISS_sum}

KISS is an H$\alpha$ emission-line-selected objective-prism survey for extragalactic 
objects \citep{s00}.  It is the first fully digital objective-prism survey for 
emission-line galaxies (ELGs), with all survey data acquired using the 0.61 m Burrell 
Schmidt telescope located on Kitt Peak.  The current study makes use of the first two 
survey strips.  The first, described in \citet{s01}, coincides with the Century Redshift 
Survey (Geller et al. 1997; Wegner et al. 2001), covering the region RA = 12h 15m to 17h 
0m, Dec = 29 - 30$^o$, while the second strip, described in \citet{kiss43}, cuts through 
the Bo\"{o}tes void (RA = 11h 55m to 16h 15m, Dec = 42.7 - 44.3$^o$).  The first survey 
strip was obtained using a 2048 $\times$ 2048 pixel STIS CCD (S2KA) which has 21 $\mu$m 
pixels with an overall field of view of 69 $\times$ 69 arcmin.  Each image covered 1.32 
deg$^2$.  The second survey strip was taken with a SiTe 2048 $\times$ 4096 CCD with a 
coverage area of 50 $\times$ 100 arcmin (giving an area of 1.39 deg$^2$).  The pixel 
size of the new CCD is smaller (15 $\mu$m), so the area covered by each image is nearly 
equivalent to the original STIS CCD.  

The H$\alpha$-selected (hereafter referred to as ``red'') portion of the KISS 
objective-prism spectral data covers a spectral range of 6400 - 7200 \AA.  The 
red survey spectral data were obtained with a 4$^{\circ}$ prism, which provided a 
reciprocal dispersion of 24 \AA\ pixel$^{-1}$ at H$\alpha$ ($\lambda_o$ = 6563 \AA) 
for the survey strip at Dec = 30$^o$, and 17 \AA\ pixel$^{-1}$ for the 43$^o$ strip.  
The spectral range was chosen to have a blue limit just below the rest wavelength
of H$\alpha$ and to extend up to the beginning of a strong atmospheric OH molecular 
band near 7200 \AA.  H$\alpha$ is detectable in this range up to a redshift of 
approximately 0.095 \citep{s01}.  Along with the objective-prism images, direct 
images (images with no prism in front of the telescope) for each field are obtained.  
Images were taken through two filters (\emph{B} and \emph{V}) to a depth of $B = 21-22$, 
which is 1 to 2 magnitudes fainter than the limiting magnitude of the spectral images.  
From these images, accurate positions and \emph{B} and \emph{V} photometry for all 
objects are obtained.

The ELG candidates selected by KISS are cataloged, but these remain only 
candidates until follow-up spectroscopy can be done for these objects.  
Higher resolution slit or fiber spectra have been obtained and reduced for 
approximately 83\% of the 30 degree strip, and 30\% of the newer 43 degree 
strip.  These spectra cover a wide range of optical wavelengths, which allow 
for the measurement of various emission lines such as H$\alpha$, H$\beta$,
[O III] $\lambda\lambda$ 4959, 5007 \AA, [N II] $\lambda\lambda$ 6548, 6583 \AA, 
etc.  From these spectra, the activity type of the ELGs can be determined 
(Seyfert 1 and 2s, starbursts, Low Ionization Nuclear Emission Regions (LINERs), 
QSOs) as well as the metallicities of these galaxies \citep{jackie}. 

\subsection{Faint Images of the Radio Sky at Twenty-centimeters -
  FIRST} \label{FIRST_sum}

The FIRST survey is being conducted with the Very Large Array (VLA) in the B
configuration at 1.4 GHz.  The stated sensitivity of the FIRST source catalog 
is 1.0 mJy/beam, with a typical rms noise level of 0.15 mJy/beam, and an angular 
resolution of 5.4 arcsec FWHM \citep{beck95}.  The sensitivity and resolution 
were selected to make FIRST comparable to the Palomar Observatory Sky Survey (POSS) 
in depth and resolution at radio frequencies, and is the highest resolution 
large-area radio survey available.  FIRST detects an average source density of 
89.5 sources per deg$^2$.  A map is produced for each field and sources are 
detected by using an elliptical Gaussian-fitting procedure \citep{white97}.  
The catalog produced from this survey was estimated to be 95\% complete at 2 
mJy/beam and 80\% complete at 1 mJy/beam \citep{beck95}. 

Not only did FIRST fully overlap the area of the sky observed by KISS, but its 
scientific priorities complemented KISS and this radio-optical matching project.  
Using the B configuration of the VLA, FIRST was able to achieve positional accuracies 
of better than 1'', which is sufficient to obtain a large number of optical 
identifications with low chance coincidence rates \citep{beck95}.   For an object 
with an optical \emph{V} magnitude of 20, the FIRST data have an identification error 
rate of 1\%, while surveys taken in the C and D configurations have error rates of 
$\sim$6\% and $\sim$25\%, respectively \citep{beck95}, when dealing with radio
point sources.  However, the small size of the beam (5.4 arcsec) can also break up 
a single extended radio source into multiple FIRST sources.  Thus, it is possible 
for an unassociated radio lobe to appear to be matched with a nearby KISS galaxy, 
while the actual central source of the radio emission is at a larger offset on the sky.  
Our method of checking for this situation is explained in Section \ref{F-K_match}.  
Another significant effect of the small beam size is that FIRST is not sensitive to
extended, low-surface-brightness emission when compared to surveys with larger beam 
sizes, such as NVSS.

FIRST is sufficiently sensitive to detect the star-forming galaxies that have
been proposed to start dominating the radio population within the flux regime 
of $\le 1$ mJy \citep{george99,pran01}.  This correlates well with KISS which is 
sensitive to starburst galaxies over a large volume.  \citet{beck95} state that 
at the FIRST threshold, $\sim$75\% of detected galaxies will be AGNs and
the remainder are star-forming galaxies.  While this may be true for the full
FIRST database, we find a very different mix of AGNs and star-forming galaxies
among the radio-detected KISS galaxies (see Section 3.4).

\subsection{NRAO VLA Sky Survey - NVSS} \label{NVSS_sum}

The NVSS is a VLA radio survey that made 1.4 GHz continuum total intensity and 
linear polarization images utilizing the compact D configuration \citep{c98}.  
The survey was sensitive down to 2.5 mJy/beam with rms brightness fluctuations of 
$\sigma \approx$ 0.45 mJy/beam and an angular resolution of 45 arcsec.  The survey 
covers 82\% of the celestial sphere, including the KISS area.  While NVSS does not
probe as deep as FIRST for point sources, the increased beam size allows NVSS to 
detect low-surface brightness extended radio sources, some of which are not detected 
by FIRST.  

NVSS takes a different approach to compiling a radio survey than does
FIRST.  The most fundamental requirement in their survey is
completeness.  FIRST will detect fainter radio objects, but
NVSS's goal is to detect all local radio emitting galaxies with
uniformly accurate flux densities.  The large beam size does not break 
up extended radio sources into multiple sources as FIRST does, so NVSS
tends to be more accurate in its flux measurements of extended sources.
However, NVSS is not as sensitive at detecting faint point sources,
which FIRST is designed to detect.

As \citet{c98} state, no single survey can combine the positive attributes of
both a low resolution survey (for low-brightness sources and accurate flux densities 
of extended sources) and a high resolution one (for accurate radio positions and 
radio-source identifications with faint galaxies).  Our solution to this problem is 
to combine the results of the radio-optical matching of FIRST and NVSS radio sources
with the KISS sample.  FIRST provides us with high positional accuracy of the radio 
sources, as well as faint point source radio detections.  Alternatively, NVSS 
provides us with very accurate flux measurements as well as sensitivity to faint 
diffuse radio emission.  The KISS project can then determine the activity type 
classification as well as other spectroscopic information, such as metal abundances 
and redshifts, by obtaining follow-up spectroscopy of the optical counterparts.

\section{Radio-Optical Matching and Sample}

The radio-optical matching procedures for FIRST-KISS and NVSS-KISS differ significantly 
due to the differing approaches, resolutions, and parameters used by the FIRST and NVSS 
surveying teams.  Both procedures utilize the KISS direct images and the Digitized Sky 
Survey (DSS) optical images as the comparison to the FIRST and NVSS radio emission. 

\subsection{FIRST-KISS} \label{F-K_match}

\begin{figure*}[htp]
\epsfxsize=5.0in
\epsscale{1.0}
\plotone{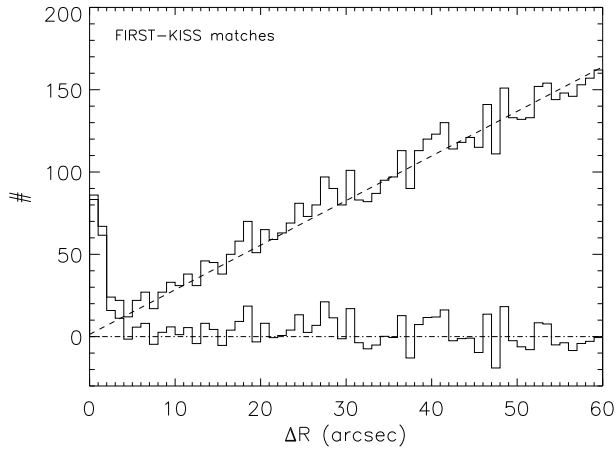}
\figcaption[plots/plot10p.eps]{FIRST-KISS positional offsets of 10 randomly
  selected fields for all objects matched out to 60 arcsec separations.
  The dashed
  line represents a linear fit with an intercept constrained to zero.
  The lower histogram is the difference between actual matches and the
  linear fit at each bin.  Matches with $\Delta$R of less than 2 arcseconds 
  have a low proportion of spurious detections. \label{fig:init_match}}
\end{figure*}

Both the FIRST survey and KISS have relatively high angular resolutions (5.4 arcsec for 
FIRST, 3--4 arcsec for KISS), which provides accurate sky positions for the objects 
contained in both surveys.  Because of this high level positional accuracy, we chose to 
perform a direct positional comparison between the surveys.  We overlaid the KISS direct 
image of each survey field with the positions of the FIRST sources covering the
KISS field of view.  Figure \ref{fig:init_match} displays the results of a positional 
matching exercise for a random sampling of ten KISS fields and includes all the FIRST 
object matches to any KISS ELG with positional offsets of under 60 arcseconds.  
The linear increase of matches at large $\Delta$R is indicative of the increase of 
spurious matches at larger and larger radii (since the areas of the annuli around 
each target galaxy increase linearly with $\Delta$R).  A linear fit through the 
origin out to 60 arcseconds divides the sample into the real and spurious matches.  
The lower histogram shows the difference between the full set of matches and the 
linear fit.  The fraction of spurious matches is very low for
separations under 2 arcseconds ($\sim$1\%).  Based on these results, we decided to 
accept any radio source match that has a $\Delta$R $< 2.0$ arcsec as a valid match 
as long as the radio source was unresolved (i.e., a point source).  Following this 
decision we visually examined the remaining putative matches.  Due to the angular 
size of these sources, we assume that real point source matches are only possible 
with offsets under 10 arcsec.  True matches of point radio sources to optical 
galaxies should be close to the central region of the galaxy.  True radio source 
matches at greater offsets should only be extended sources, i.e, lobe sources
from an AGN or extended radio emission from a starburst region of large angular size.
The remaining radio source matches with $\Delta$R $ > 2$ arcsec required visual 
verification (see below) but very few were accepted as $bona fide$ matches.  We 
did not attempt to match radio sources at $\Delta$R $ > 30$ arcsec due to the high 
resolution and precision of KISS and FIRST source positions and the increasing number 
density of point sources at these radii, which will only contribute spurious sources 
to our radio-optical sample.  

When carrying out our matching procedure, we were careful not to exclude possible 
matches of radio sources with extended lobe morphologies with optical 
counterparts at larger $\Delta$R.  Thus, we chose any KISS object within 30 arcsec 
of an {\it extended} FIRST source to be a potential match to that source.  The radio
positions and morphologies of these sources were visually checked by utilizing 
the FIRST survey field cutout web tool (http://sundog.stsci.edu/), which we used 
extensively to reduce the identification error rate even further.  By utilizing the 
KISS object position, we can determine if the radio source is sufficiently centered 
in the FIRST radio field image.  This is enough for verifying point sources with 
$\Delta$R $> 2.0$ arcsec, but the radio field image also allows us to determine if an 
extended source is a radio lobe that has its emission source centered on the KISS 
object.  In the end, no such lobe sources were accepted as positive matches.

\begin{figure*}[htp]
\epsfxsize=3.0in
\epsscale{2.0}
\plottwo{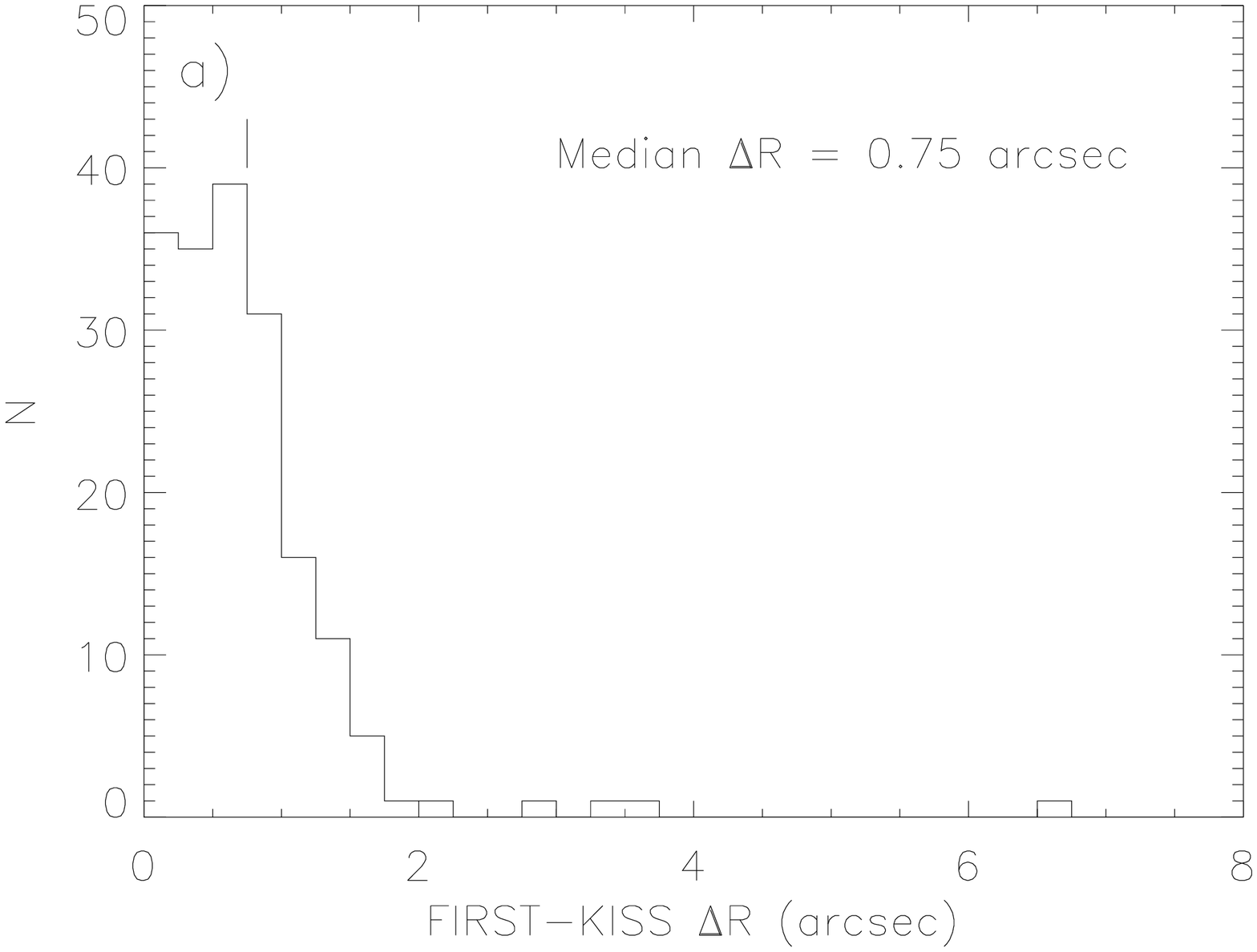}{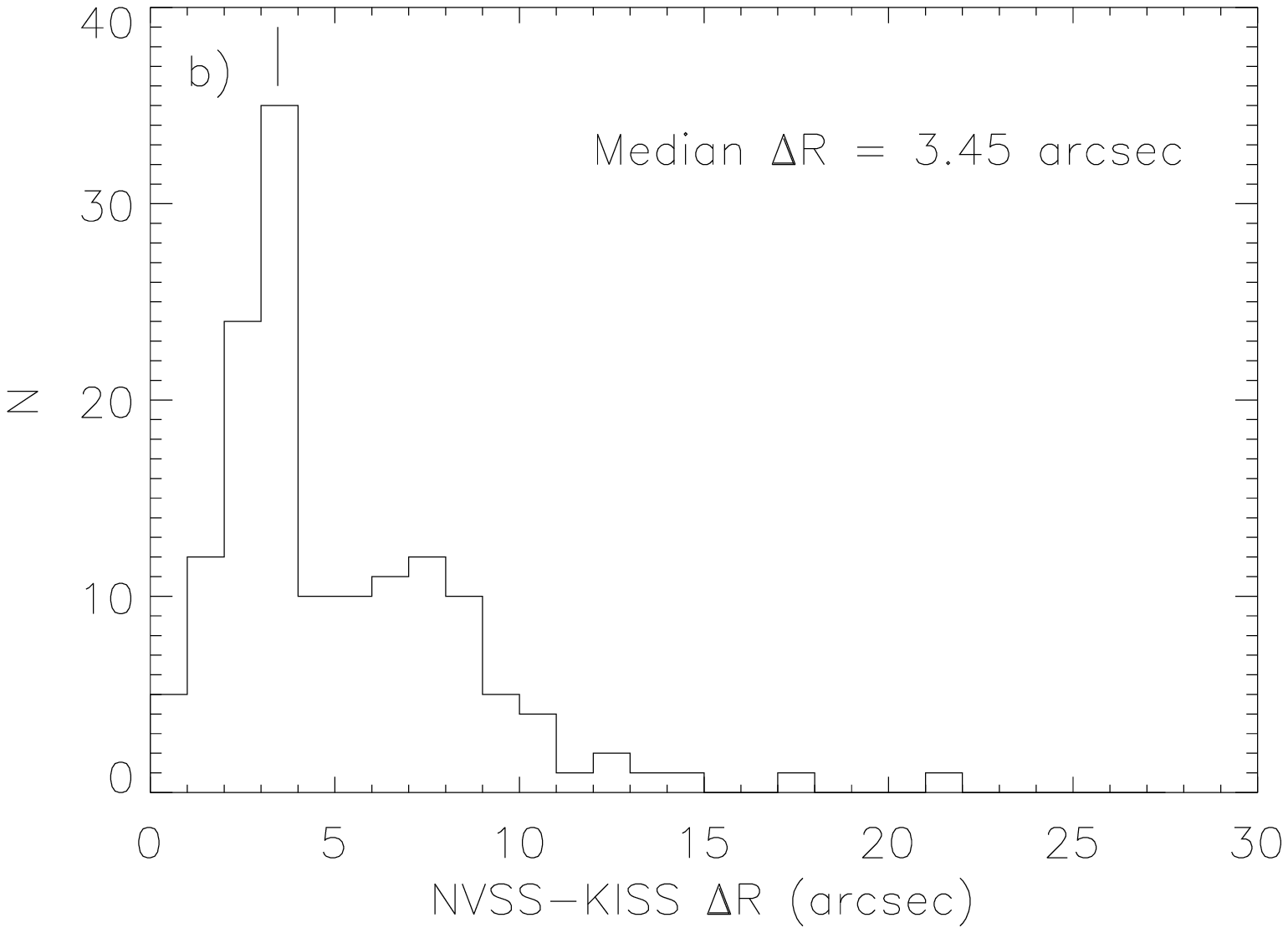}
\figcaption[plots/firstdelrf.eps]{Distribution of the (a) FIRST-KISS and
  (b) NVSS-KISS positional offsets of verified source matches. \label{fig:delrf}}
\end{figure*}

Out of a possible 2157 KISS ELGs, 184 FIRST radio matches were verified.  Figure 
\ref{fig:delrf}a shows the distribution of positional offsets for the matched ELGs.  
The few matches beyond 2 arcsec were visually verified.  The median value of 
the offsets (0.75 arcsec) further strengthens our confidence that $\sim 99\%$ of the 
matches within 2 arcsec are real matches.

\subsection{NVSS-KISS}

The 45 arcsec angular resolution of NVSS requires a different approach
to the positional matching of the KISS galaxies.  Using Digitized Sky Survey 
(DSS) images of the KISS field of view, we overlaid NVSS radio contours to aid 
our search for true positional matches.  Finder fields of every KISS object with
an NVSS radio source within 60 arcsec were generated.  Using the
intensity contour lines as guides, we found that the majority of NVSS
matches were visually centered on the optical KISS object.  However,
their positional offsets were typically much higher than the FIRST
offsets shown in Figure \ref{fig:delrf}a.  Those radio contours that were
not directly centered on an object were examined to a fainter contour
level for a KISS object, unless the morphology suggested a more
complex lobe structure.  Furthermore, for any source match that we could
not decide upon by eye, we input the radio position into the FIRST radio 
field cutout tool.  The FIRST images usually had at least some radio emission 
above the noise (2 - 3$\sigma$), sufficient to decide if the radio source was 
indeed associated with the matched KISS object.

The NVSS-KISS comparison found 23 matches that the FIRST-KISS comparison did not, 
all of them extended radio sources.  In addition, there are 124 NVSS-KISS matches 
that were also found by the FIRST-KISS comparison.  An additional 60 sources
were detected only by FIRST.  By performing the FIRST matches 
prior to the NVSS matches, we knew which objects were matched to FIRST radio
sources.  We then have more confidence in our verification of complex or weak 
NVSS contour lines around a KISS object. Figure \ref{fig:delrf}b displays $\Delta$R
for all NVSS matched sources.  A high fraction of matches (82\%) are found within 
8 arcsec, and the median $\Delta$R is 3.45 arcsec.  This low median value increased 
our confidence in our positional accuracy.  In total, 207 KISS ELGs were matched to 
radio sources.

\subsection{Radio-flux adjustments}

\subsubsection {Improving flux measurements}

FIRST uses a two-dimensional Gaussian profile to measure the flux of all FIRST 
radio sources.  This method is accurate for point sources, but the use of this fixed 
function will underestimate the flux of extended FIRST sources, especially radio 
lobe sources.  To rectify this problem, we downloaded the public domain FIRST 
FITS files from the FIRST website for all apparently extended FIRST sources, and 
measured their total (integrated) radio fluxes using the IRAF PHOT task.   Using 
the IRAF photometry of 28 point source radio galaxies with FIRST integrated fluxes 
ranging from 0.8 mJy to 402.0 mJy as calibration sources, we derived the following 
relationship:
\begin{equation}
F_{IRAF}= (10.185 \pm 0.002) F_{int}
\end{equation}
where $F_{IRAF}$ is the flux measured by IRAF in counts, and $F_{int}$ is the 
integrated FIRST flux in mJy.  This relationship was used together with our 
IRAF-based measurements of the total radio fluxes to arrive at revised estimates 
of F$_{int}$ for the extended FIRST objects.  Of the 19 candidates for re-measurement, 
9 had at least a 10\% increase in flux, with one object having a flux increase from 
3.62 mJy to 242.3 mJy after including flux from associated radio lobes.  Only those 
measurements with an increase of at least 10\% had their fluxes adjusted.  Table 
\ref{tab:fluxadjust} contains the original and re-measured flux values.


\subsubsection{Comparison of FIRST and NVSS flux scales}

\begin{figure*}[htp]
\epsfxsize=4.0in
\epsscale{1.5}
\plotone{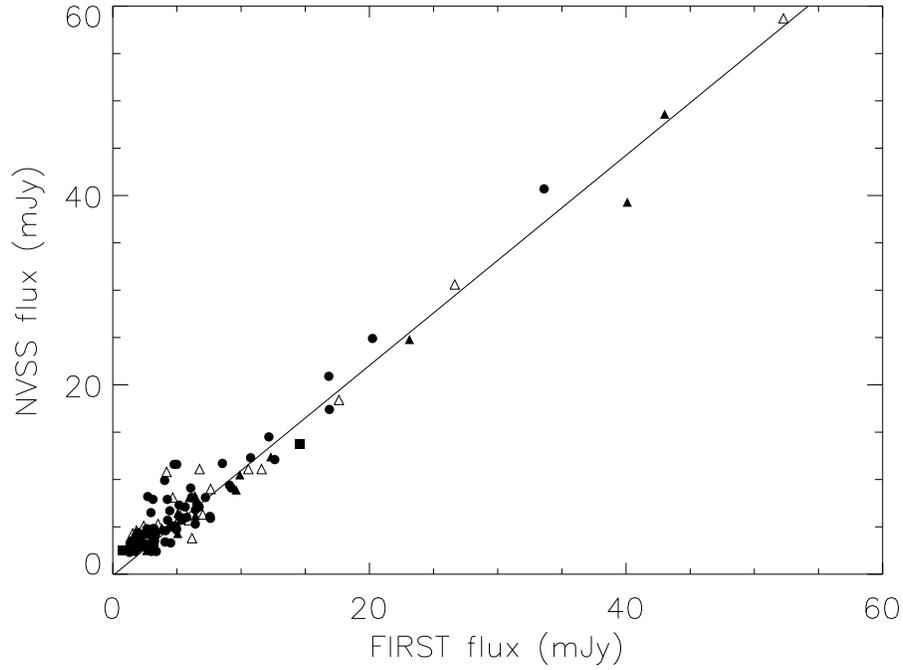}
\figcaption[plots/firstnvsscomp.eps]{Flux density comparison between
  galaxies detected by both FIRST and NVSS.  The least squares linear fit (see
  Eq.\ref{eq:firstnvss}) is used to rectify the different flux scales.  Symbols 
  represent different ELG types: Seyfert 1s (squares), Seyfert 2s (filled triangle),
  starbursts (circles), and LINERs (open triangle).  \label{fig:surveycomp}}
\end{figure*}

\begin{figure*}[htp]
\epsfxsize=4.0in
\epsscale{1.5}
\plotone{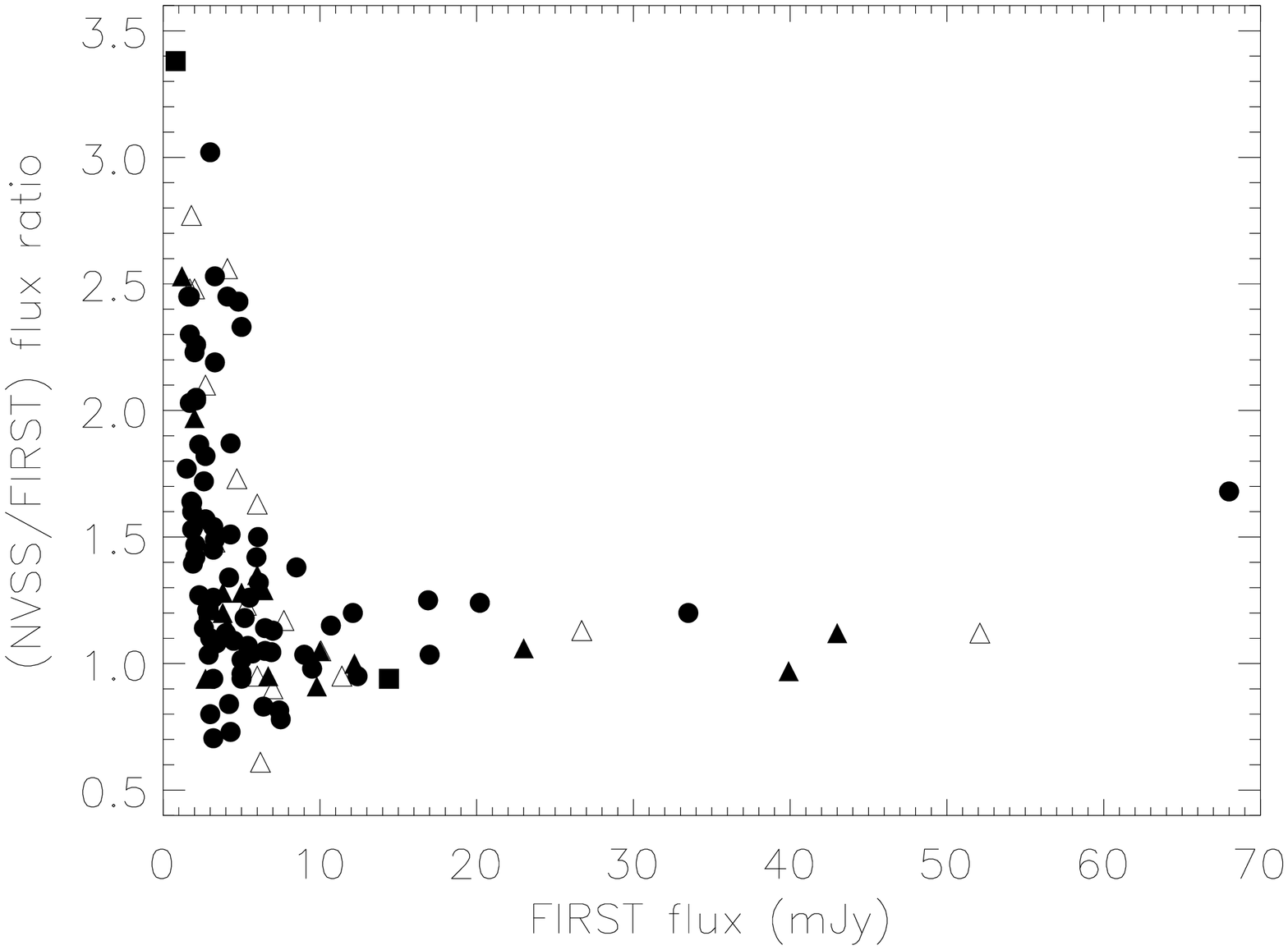}
\figcaption[plots/firstnvssratio.eps]{The ratio of the NVSS to FIRST radio 
  fluxes vs. the FIRST survey flux for the same galaxies plotted in Figure 
  \ref{fig:surveycomp}.  The mean ratio for sources with F$_{FIRST}$ between 
  5 and 60 mJy is 1.10. \label{fig:surveyratio}}
\end{figure*}

In order to combine the matching results from the FIRST and NVSS
samples, we had to determine the relationship between the flux scales
of the two surveys.  We compiled a list of all 124 KISS ELG matches
that had detections by both FIRST and NVSS.  We plot the NVSS flux
{\it versus} FIRST flux in Figure \ref{fig:surveycomp}, 
as well as the NVSS/FIRST flux ratio {\it versus} the FIRST flux in Figure
\ref{fig:surveyratio}.  The flux ratio plot shows a large discrepancy 
between measurements of radio sources with FIRST fluxes less than 5 mJy.  
This is not unexpected, as extended, low flux galaxies detected in both 
surveys will be resolved out by FIRST.  In order to quantify the differences 
between the FIRST and NVSS fluxes, we performed a number of statistical comparisons.  
In doing so, we constrain the calculations to radio sources 
with the flux range of $5\le F_{FIRST} \le 60$ mJy.  This is because at lower 
fluxes we get much larger discrepancies in flux primarily due to the difference 
in beam size of the two surveys.  At fluxes within our chosen range, which are 
more likely point sources, the scaling is far more consistent. The upper limit 
was chosen to avoid small-number statistical effects and because the high flux 
objects tend to show extreme discrepancies between FIRST and NVSS.

The line shown in Figure \ref{fig:surveycomp} is determined by a linear least 
squares fit to the galaxies in the flux density range defined above.  The fit
yields the equation:
\begin{equation}
F_{NVSS} = (1.110\pm0.024) F_{FIRST} - (0.152\pm0.394).
\label{eq:firstnvss}
\end{equation} 
The mean value of the NVSS/FIRST flux ratio using the same galaxies is found to 
be $\mu_{ratio}=1.099\pm0.028$, with a median of 1.061 and a standard deviation 
of $\sigma = 0.194$.  The slope of the linear fit and the mean ratio value agree to 
within the errors, as they should.

For consistency throughout the radio-KISS sample, we chose to adjust
the NVSS fluxes to the FIRST fluxes by dividing all NVSS-only
detected radio sources by 1.10.  Any galaxy that was matched to both
FIRST and NVSS retained the FIRST flux value.  This is not to imply that 
the NVSS fluxes are inferior to the FIRST fluxes.  On the contrary, they 
appear to be more accurate, in the sense that they include more of the 
extended low surface brightness emission than does FIRST.  However, since 
FIRST sources outnumber NVSS detections, we decided to scale the fluxes to 
the FIRST scale for subsequent analysis.

\subsection{Final Sample}\label{finalsample}

In Table~\ref{firstnvsstypes}, we present the number of KISS galaxies
with FIRST and NVSS radio source matches, broken down by ELG type.  A 
total of 207 KISS ELGs out of the entire KISS sample of 2157 ELG candidates 
were successfully matched with a radio source from either FIRST or NVSS.  
FIRST-KISS matched 184 galaxies while NVSS-KISS matched 147 galaxies.  A 
total of 60 galaxies were matched only by the FIRST survey (all low flux 
sources) and 23 galaxies were matched only by NVSS (mostly slightly extended 
radio sources with lower radio surface brightness than the rest of the 
sample).  In the cases that both surveys have sources matched to the same 
KISS object, the FIRST flux has been used in the analysis.  All 207 radio-detected
KISS galaxies possess follow-up spectra.  Note that 10 galaxies are actually high 
redshift (i.e. $z > 0.3$) objects and do not lie within the H$\alpha$ selection
volume (to z = 0.095).  These ELGs were detected by KISS via their [O III] 
$\lambda\lambda$4959, 5007 \AA\ lines, which had been redshifted into the wavelength 
range used by KISS.  One exception is the QSO KISSR 844, which is at a redshift of
z=0.46 and was detected by H$\beta$.  In order of contribution to the radio sample, 
the activity types of the radio-detected ELGs are: starburst/star-forming galaxies 
(63.8\%), LINERs (17.9\%), Seyfert 2s (15.5\%), Seyfert 1s (2.4\%), and QSOs (0.5\%).  
The AGNs (LINERs, Seyferts, and QSOs) contribute a total of 36.2\% of the radio-detected 
KISS galaxies, which is substantially higher than the 14.7\% AGN fraction in the
overall KISS sample among those objects with existing follow-up spectra.


While only a relatively small fraction of the KISS star-forming galaxies have
radio detections (132 of 960, or 14\%), roughly half of Seyfert 2s and LINERs in 
the KISS sample that have been identified via follow-up spectroscopy have also 
been detected as radio sources.  Of the 64 Seyfert 2 galaxies identified by KISS, 
32 (50\%) are detected radio sources.  Similarly, of the 76 identified LINERs, 37 
(49\%) have radio emission.   The large number of LINERs in relation to the overall 
sample of radio emitting ELGs is unusual and surprising.  Previous deep radio surveys 
have typically found only very few LINERs.  Furthermore, previous optical 
objective-prism surveys, traditionally carried out in the blue portion of the spectrum, 
have missed most galaxies we would classify as LINERs.  The combination of the 
H$\alpha$-selection method used by KISS and the excellent sensitivity of the two radio 
surveys is presumably the reason for the high number of LINERs present in the current 
sample.  Somewhat surprisingly, only 5 of the 20 identified Seyfert 1s (25\%) have 
associated radio emission.  Note that the numbers above include the high redshift
objects.  If these are excluded the disparity between the two Seyfert classes becomes 
even worse.  For the Seyferts with z $<$ 0.08, 2 of 11 Seyfert 1s (18\%) are radio 
detected, as opposed to 21 of 35 Seyfert 2s (60\%).  

A {\it naive} interpretation of the unified model for AGN (e.g., Schmidt \etal 2001; 
Meier 2002; Veilleux 2003) would suggest that Seyfert 1s and 2s would be detected at a 
similar rate in the radio continuum, unless beaming effects are important, in which case 
the Seyfert 1s would be expected to be stronger.  Yet the Seyfert 2s greatly outnumber 
the broad-lined Seyferts in terms of radio detections.  
The apparent disparity between the fraction of radio detections in the two types of
Seyferts might be explained if some or all of the radio emission detected in the Seyfert 2s
were due to circumnuclear star-formation activity rather than being associated with the
AGN.  A test of this hypothesis could be achieved by evaluating the spatial distribution
of the radio emission: extended radio emission would indicate a star-forming origin, while
a radio point source would be consistent with pure AGN emission.  Evaluation of the radio
maps obtained by FIRST indicate that all of the KISS Seyfert galaxies detected in the radio
are in fact point sources.  None show extended emission.  However, given the typical
distance of the KISS Seyferts and the spatial resolution of the FIRST data, this is not
a sensitive test.  For example, the 5.4 arcsec FWHM for a FIRST point source corresponds
to a linear size of 7.3 kpc for a galaxy with z = 0.07.  In nearly all cases, any putative 
circumnuclear star-forming region would 
not be resolved from the central AGN in the FIRST radio maps.  Hence, we cannot resolve 
the question of whether star-formation activity is enhancing the radio emission from some
of the Seyfert 2s.  It should also be noted that there is no obvious reason why 
circumnuclear star-formation should preferentially be favored in the Seyfert 2s over
the Seyfert 1s; the unified model does not predict any such difference.  Clearly, 
further work in this area will be necessary.

It is worth noting that the percentages quoted in the preceding paragraph refer to
the fraction of KISS ELGs that currently have follow-up spectra.  While 100\% of
the radio-detected objects have follow-up spectra (i.e., they were targeted for
observation), only 58\% of the entire sample of 2157 KISS ELGs from the two survey
strips studied have been observed spectroscopically.  Hence, the percentages
of radio-detected KISS galaxies of all activity types will go down as the
follow-up spectroscopy becomes more complete.

It should also be noted that since we are using an optical line-selected galaxy 
catalog, we do $not$ detect any early-type, elliptical galaxies.  This is in 
contrast to previous studies of radio-selected samples where early-type galaxies 
are the dominant radio-emitting galaxy type at this flux level
\citep{pran01,george99,mag00,grupp99}.  This is discussed and addressed
further in \S5 and \S6.


Table~\ref{maintable} presents the primary optical and radio characteristics of 
all 207 radio ELGs found in this study.  The columns are as follows: 
(1) $KISSR$, the KISS ID of the galaxy as listed in \citet{s01} and \citet{kiss43}; 
(2) $RA$ and (3) $Dec$, given for equinox J2000.  These are the optical positions tabulated
in the survey papers \citep{s01,kiss43}, and have precisions of $\sim$0.25--0.30 arcsec;
(4) $B$: apparent magnitude in the Harris $B$ filter \citep{s01,kiss43};
(5) $M_B$: absolute $B$ magnitude (see below);
(6) $L_{H\alpha}$: H$\alpha$ line luminosity in erg/s;
(7) $S_{1.4}$: radio flux density at 1.4GHz in mJy.  If the galaxy was
    detected in both FIRST and NVSS, the FIRST flux value is listed.
    If only detected by NVSS, the flux is corrected as described above;
(8) $P_{1.4}$: radio luminosity at 1.4 GHz in W/Hz.  The luminosity
    is calculated from the distances derived from the listed redshift value
    (See \S~\ref{lumcalc});
(9) $z$: spectroscopic redshift (corrected for Local Group motion) obtained from
    the follow-up spectra;
(10) $\Delta$$R$: the positional offset of the optical-radio matches in
    arcseconds.  This value is calculated from the angular distance between the 
    peak optical emission of the KISS object and the peak radio emission;  
(11) Survey match:  indicates which radio survey the KISS object was
    matched with: F for FIRST only, N for NVSS only and B for both surveys; 
(12) ELG type: the activity type of the ELG as determined from follow-up spectra.
The ELG activity types are as follows; Sy1: Seyfert 1; Sy2: Seyfert 2; SB: starburst 
or star-forming galaxy; LIN: Low Ionization Nuclear Emission Region (LINER); and QSOs.

\subsubsection{Luminosity calculations\label{lumcalc}}

\begin{figure*}[htp]
\epsfxsize=2.5in
\epsscale{1.8}
\plottwo{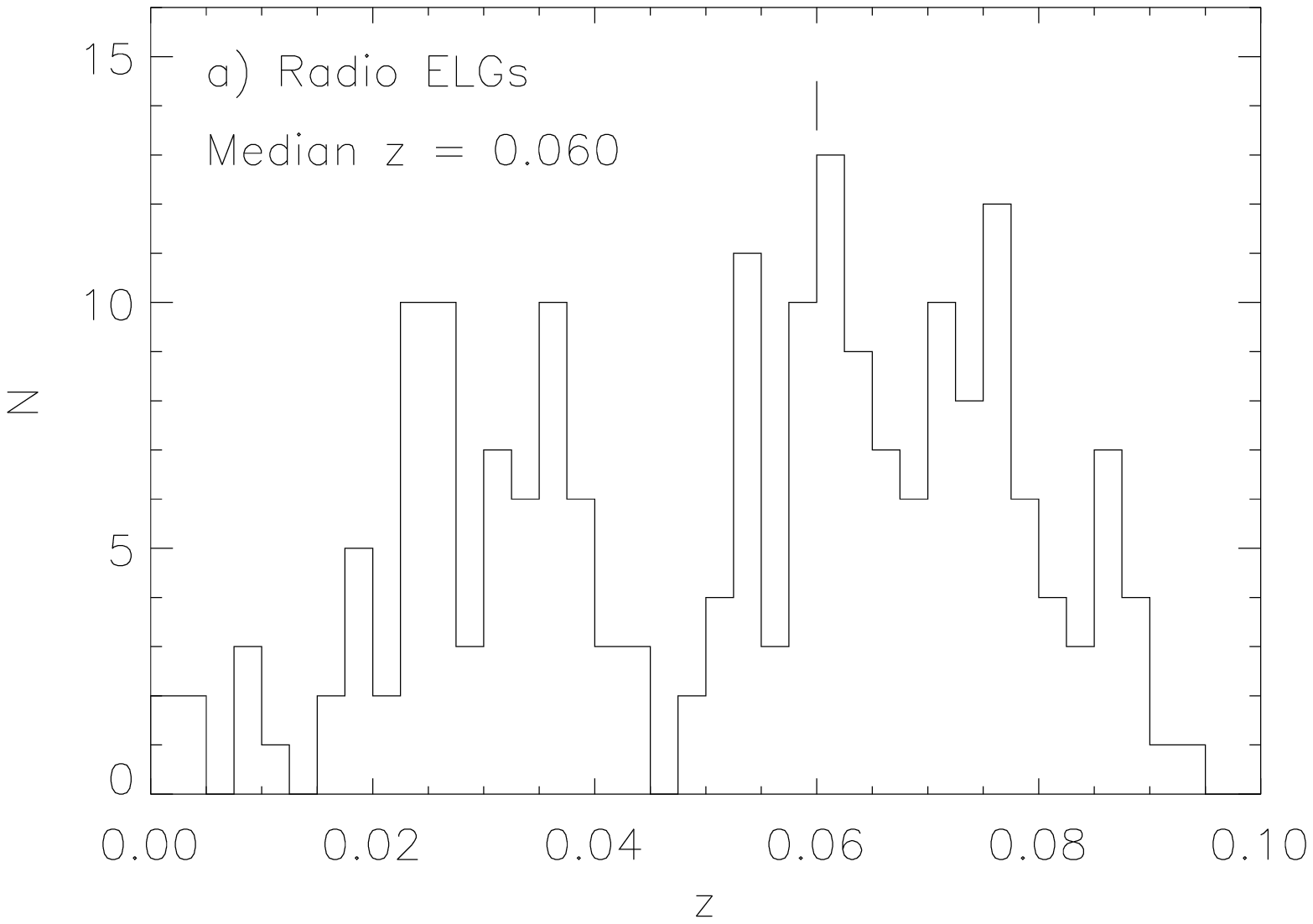}{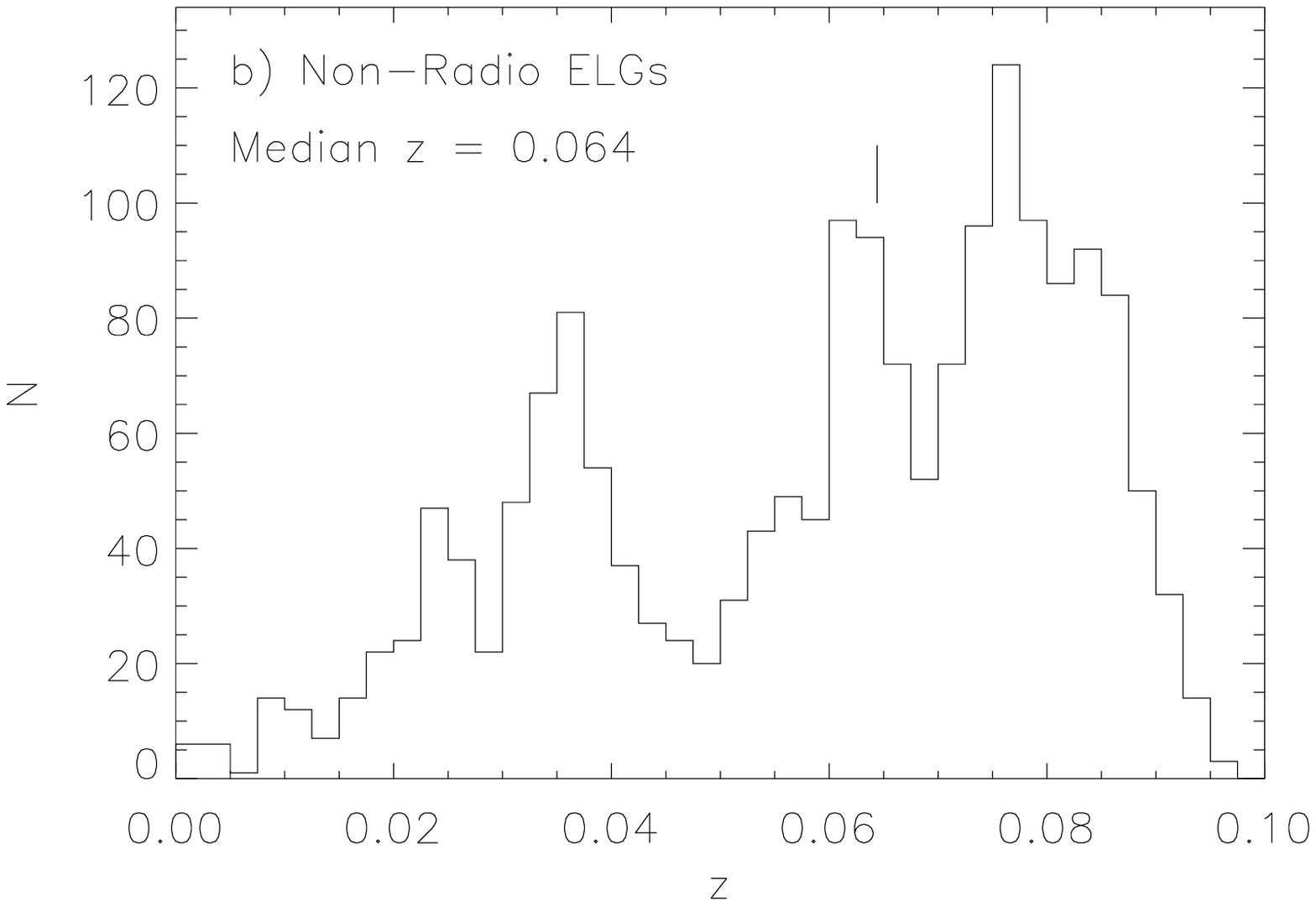}
\figcaption[plots/radio_z.eps]{Redshift distribution of (a) radio and
  (b) non-radio detected ELGs in the KISS sample.  The bimodal distribution
  is due to the low-density regions at roughly $z=0.045$ in both survey strips.
  The median redshift is essentially the same for both samples, and the two
  redshift distributions are statistically indistinguishable.\label{fig:z}}
\end{figure*}

The absolute magnitudes and luminosities in Table~\ref{maintable} were calculated 
by using the redshifts obtained through follow-up spectroscopy.   
The typical error for our spectroscopic redshifts is $\pm 30$ km/s.  
For all distance-dependent quantities, we assume H$_o$ = 75 km/s/Mpc and 
q$_o$ = 0.5.  Distances were derived from the observed redshifts, assuming 
a pure Hubble flow and after correcting for the motion of the sun relative to
the center of mass of the Local Group (v$_{corr}$ = v$_{helio}$ + 300 $\sin~l \cos~b$).
The H$\alpha$ luminosities were calculated from the H$\alpha$ flux measured from the  
objective-prism spectra \citep{s01}.  This was done to remove slit/fiber effects which 
lead to the potential for loss of flux with some of our follow-up higher resolution 
spectroscopy.   Use of the objective-prism H$\alpha$ fluxes provides a uniform,
homogeneous data set to work with, but requires us to correct for the additional
flux contributed by the nearby [N II]($\lambda\lambda$ 6583, 6548 \AA) lines that 
are blended with H$\alpha$ in the objective-prism spectra.  Additionally, we correct 
for intrinsic absorption due to dust using the Balmer decrement coefficient measured
from follow-up spectra.

\section{Radio Properties of the KISS ELGs}

In this section, we present a breakdown and comparison of the optical characteristics 
between the radio and non-radio detected KISS ELGs, as well as the radio properties 
of the radio-detected sample.  We also explore the metallicity and line ratio
characteristics of the two samples in an effort to better understand what physical 
differences lie between radio and non-radio emitting ELGs.

\subsection{Comparison Between Radio and Non-Radio ELGs}

The KISS data set provides us with a tremendous opportunity to study in detail the 
physical characteristics of radio-emitting galaxies that possess strong emission lines 
and compare them to the remainder of the KISS sample that was not radio detected.  While 
obtaining redshifts and distances to galaxies selected in optical emission-line surveys is 
a trivial matter, it is much more difficult for objects discovered in radio surveys.
Only through obtaining spectroscopy of the radio sources' optical counterpart can accurate 
distances be obtained.  However, these optical counterparts are typically extremely faint, 
when detected at all.  In the current study we resolve this problem by considering the
radio characteristics of a deep sample of optically-selected ELGs (with its built-in
redshift limit) rather than the optical characteristics of a radio flux-limited sample.  
In this case, the optical matches are all relatively bright, most with B magnitudes 18.5 
or brighter.

Figure \ref{fig:z} shows the redshift distribution of the (a) radio and (b) non-radio ELGs.   
The decrease in the numbers of galaxies at $z\approx 0.045$ is due to the Bo\"{o}tes 
void that is present within the 43 degree strip, and a corresponding low-density region 
at approximately the same distance in the 30 degree strip.  The redshift distribution of
the non-radio-detected sources (Figure 5b) increases steadily out to z $\approx$ 0.075,
after which it falls off to zero at z $\approx$ 0.095.  This cut-off in the sensitivity
of the survey is caused by the filter employed in the objective-prism observations.
At redshifts above z $\approx$ 0.075, the H$\alpha$ line starts to redshift out of the
filter bandpass (Salzer et al. 2000).  Those ELGs with the strongest H$\alpha$ emission 
lines are preferentially detected at the higher redshifts as the filter will cut off 
weaker emission.  This explains the drop off at $z > 0.075$ as a completeness effect.  
Below this redshift the survey is highly complete, but at higher redshifts it 
misses some percentage of weaker-lined ELGs.  Not shown in Figure \ref{fig:z} are the 
approximately 2\% of all ELGs in the KISS survey that are located at $z > 0.3$.  These 
objects are detected by some line other than H$\alpha$ in the bandpass of the objective-prism
spectra (typically [O III]$\lambda$5007, H$\beta$, or H$\gamma$).

The two redshift histograms appear to be quite similar, as are the median values of the 
redshift distributions ($z=0.060$ and $z=0.064$ for radio and non-radio, respectively).  
We might infer that the two sets of redshifts are drawn from the same parent population. 
Applying a KS test to the data confirms this suspicion at the 99\% confidence level: the 
radio and non-radio ELGs exhibit the same redshift distributions.

\begin{figure*}[htp]
\epsfxsize=2.5in
\epsscale{1.8}
\plottwo{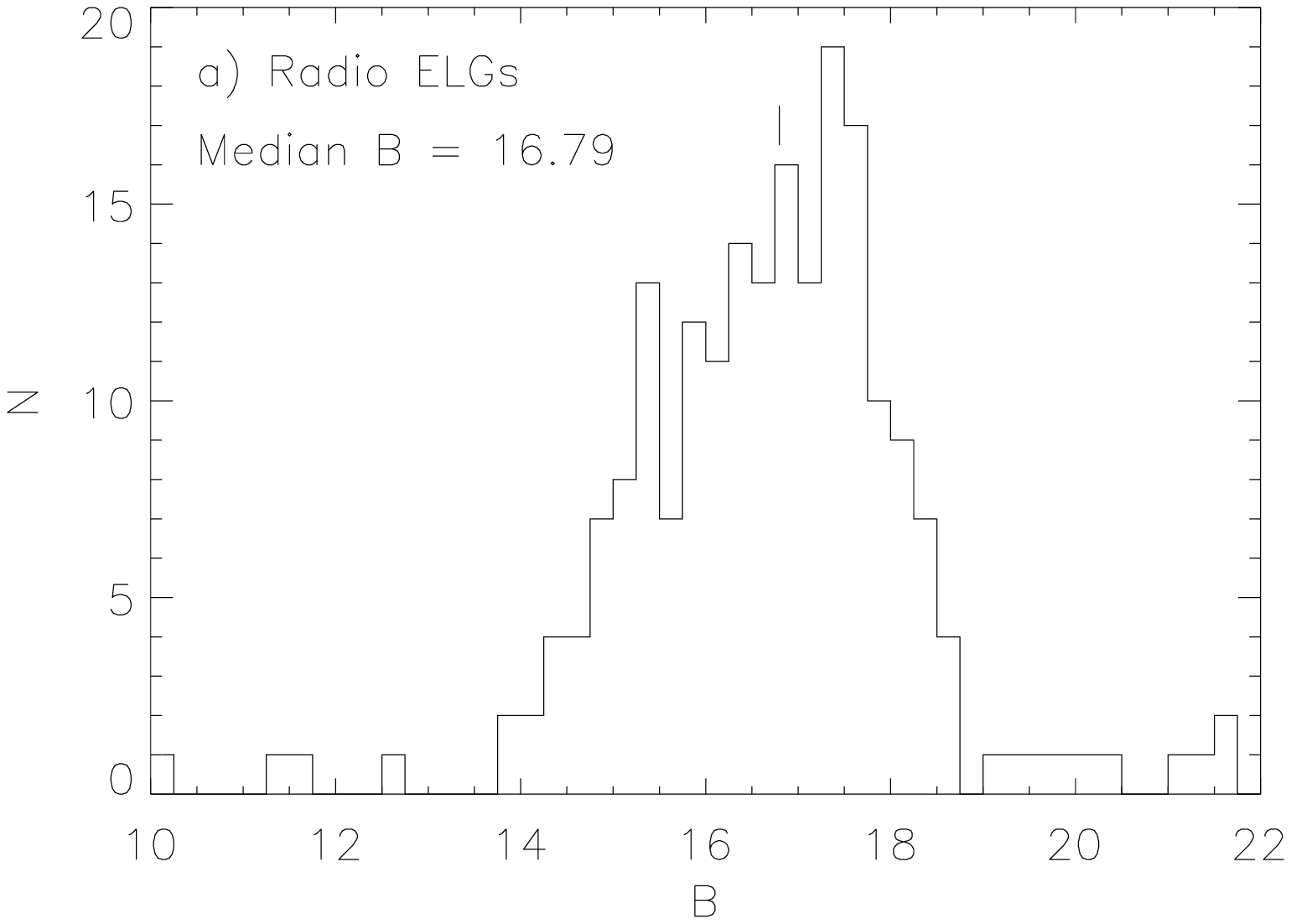}{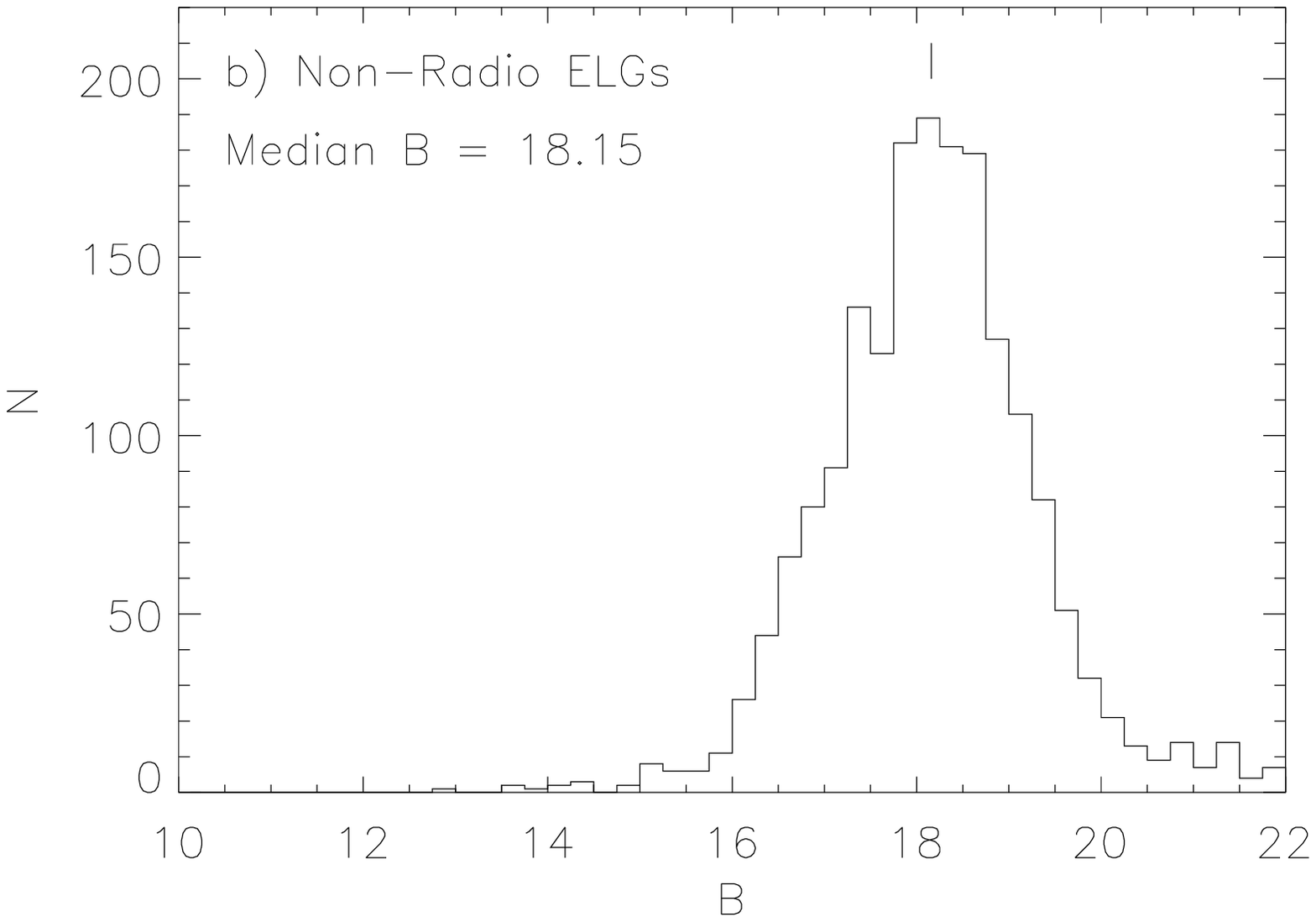}
\figcaption[plots/radio_bmag.eps]{Apparent B magnitude of (a) radio and
  (b) non-radio detected ELGs.  The radio ELGs contain the brighter
  portion of the overall KISS population and have a median 1.35 mag
  brighter than the remaining ELGs. \label{fig:bmag}}
\end{figure*}

The apparent magnitude distribution (Figure  \ref{fig:bmag}) shows that the radio ELGs
are brighter objects (median = 16.80) than ELGs without radio emission (median = 18.22).
This is a real effect, as KISS is sensitive to faint dwarf ELGs, shown by the tail of 
counts at $B\ge20$ in Figure \ref{fig:bmag}b.  Few galaxies with $B > 18$ are radio detected, 
and those that are tend to be the high redshift objects ($z > 0.3$) mentioned above.
Approximately half of all ELGs brighter than $B=16$ are radio detected.  In contrast, the 
typical optical counterparts to faint radio sources tend to be very faint.  The reason the
KISS radio sources are so bright is due to the redshift limit imposed by the survey filter
combined with the fact that radio-emitting galaxies tend to be fairly luminous (see below).  

\begin{figure*}[htp]
\epsfxsize=2.5in
\epsscale{1.8}
\plottwo{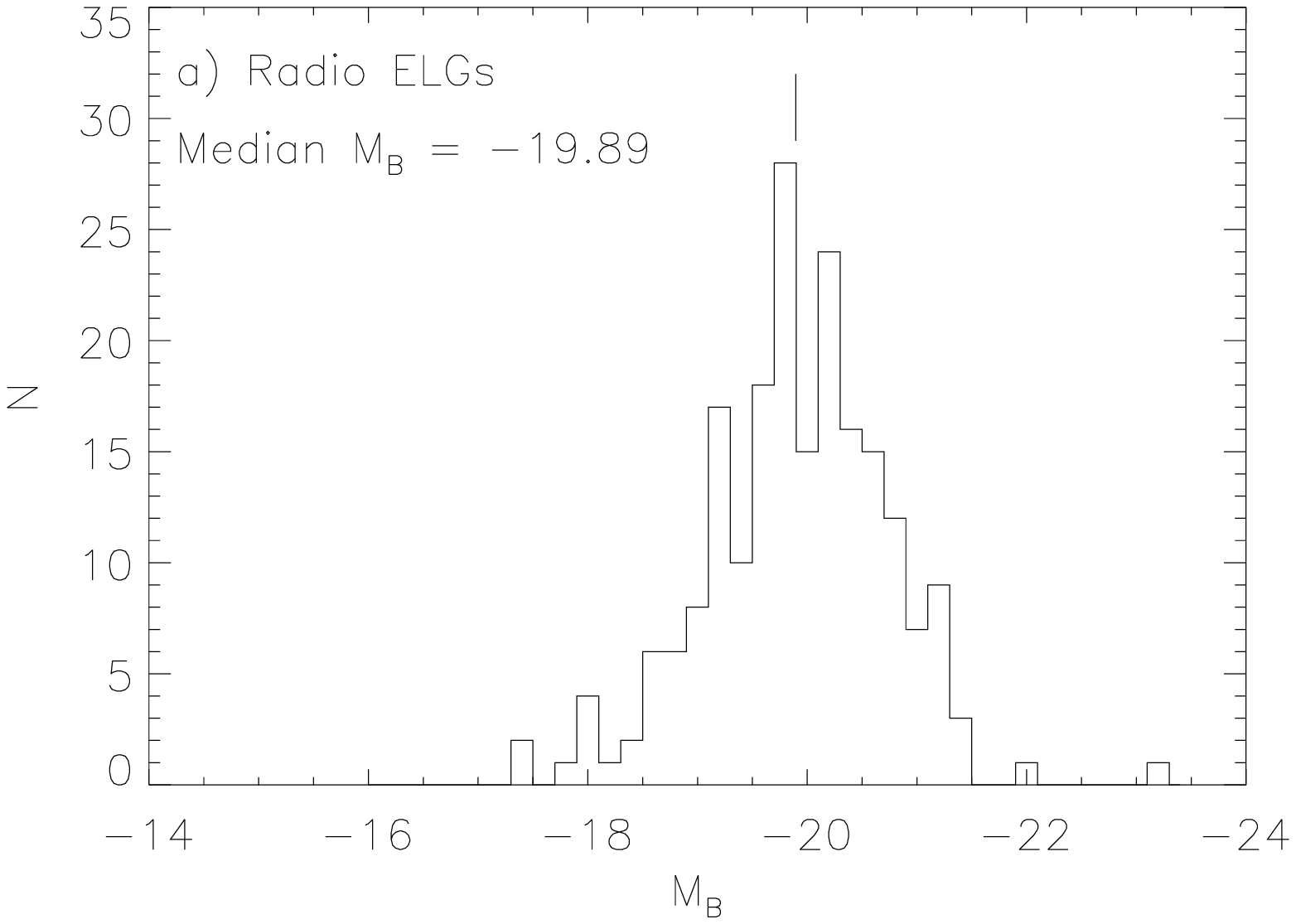}{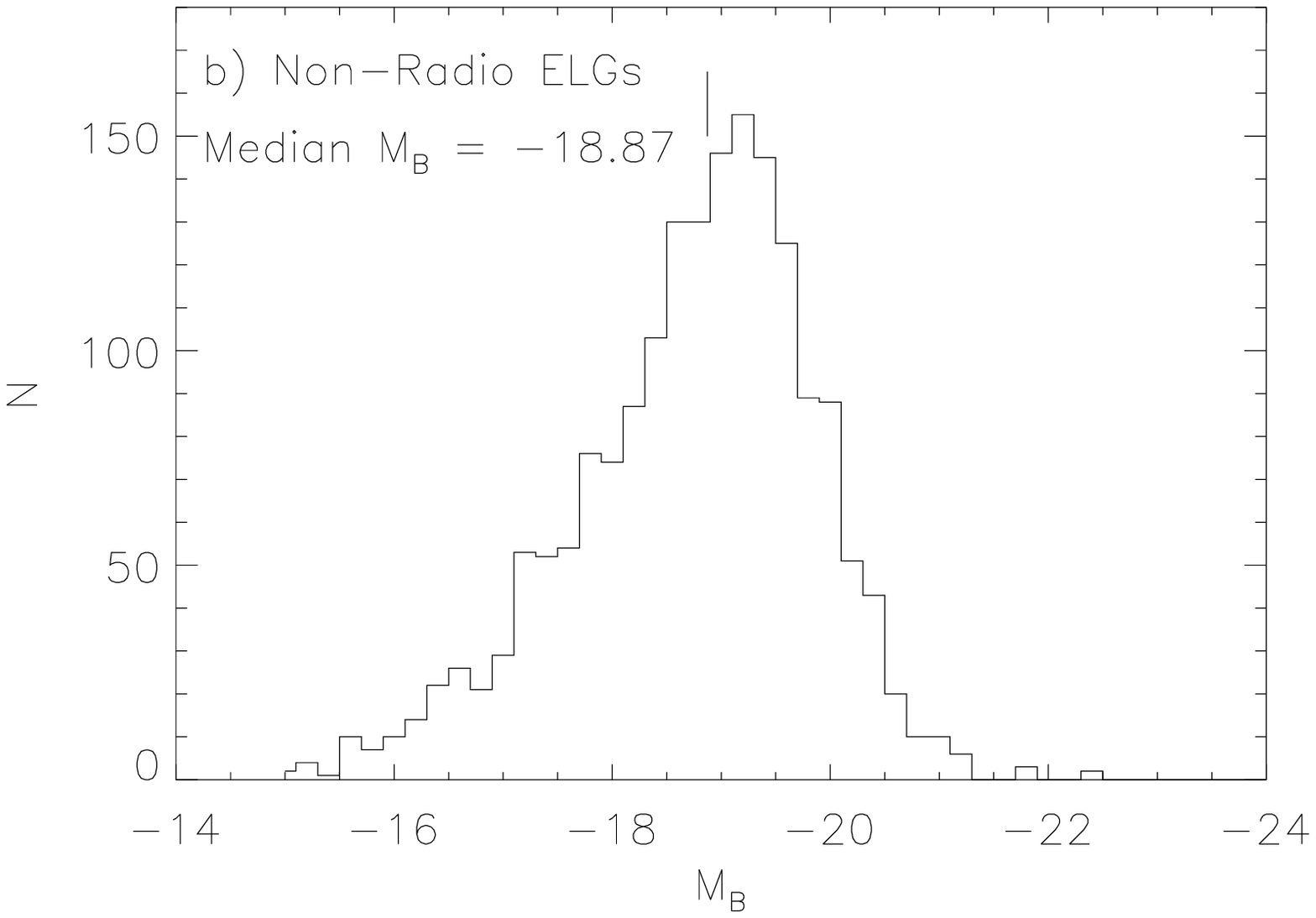}
\figcaption[plots/radio_absmag.eps]{Absolute B magnitude comparisons.
  The radio ELG population makes up a large percentage of the optically
  most luminous portion of the overall KISS sample.  No low luminosity
  galaxies (i.e. dwarf galaxies) of $M_B > -17$ have radio detections. 
  \label{fig:absmag}}
\end{figure*}

A better representation of the physical characteristics of the galaxies that are being 
detected in the radio is found in Figure \ref{fig:absmag}, which plots histograms of
the absolute magnitude M$_B$.  There are two particularly interesting features of note.  
First, the radio population is much more luminous than the non-radio.  Figure \ref{fig:absmag}a 
shows that the radio-detected ELGs display a relatively tight distribution of absolute B 
magnitudes with a median value of -19.89 mag.  The median M$_B$ is 1.02 mag brighter than for
the non-radio KISS ELGs.  The more luminous, and hence, more massive galaxies of the sample 
are the radio emitters \citep{c92}.  However, there is a sharp cut-off at the high luminosity 
end.  There are very few galaxies more luminous than M$_B$ = $-$21, and most of these are high 
redshift objects (z $>$ 0.3).  Thus, no exceptionally luminous galaxies have been radio 
detected within the H$\alpha$-selected KISS survey volume.  

Second, Figure \ref{fig:absmag}b shows a low-luminosity tail at $M_B\ge -17$, which represents 
the many dwarf ELGs that KISS detects within our volume.  However, within the radio distribution, 
there are {\it no} galaxies fainter than M$_B > -17$, and only a few with M$_B > -18$.  
To emphasize the lack of galaxies with low optical luminosities within the radio-detected sample,
consider that a galaxy with $P_{1.4GHz} = 10^{21}$ W/Hz would be detectable out to a redshift
of z = 0.023 by FIRST.  There are 130 KISS ELGs with z $\le$ 0.023.  Of these, 42 have M$_B$ 
brighter than $-$18, and 15 of these (35.7\%) are radio detected.  The remaining 88 have M$_B$
fainter than $-$18, and only two are radio detected (2.3\%).  If the low-luminosity ELGs were
detected in the same proportion as the higher-luminosity sample, there should be 31 radio 
detections in the former sample.  If the low-luminosity ELGs, which are all star-forming objects, 
were detected at the same percentage as the overall population of KISS star-forming galaxies 
(13.75\%, see section 3.4), there would still be 12 with radio detections.  A possible
implication for the low proportion of dwarf galaxies within the KISS radio sample is that these 
low luminosity, low mass galaxies are not massive enough to produce a galactic-scale magnetic 
field of sufficient strength to confine the starburst's relativistic charged particles that give 
rise to radio synchrotron emission.  This point is discussed further in Section 4.2

\begin{figure*}[htp]
\epsfxsize=2.5in
\epsscale{1.8}
\plottwo{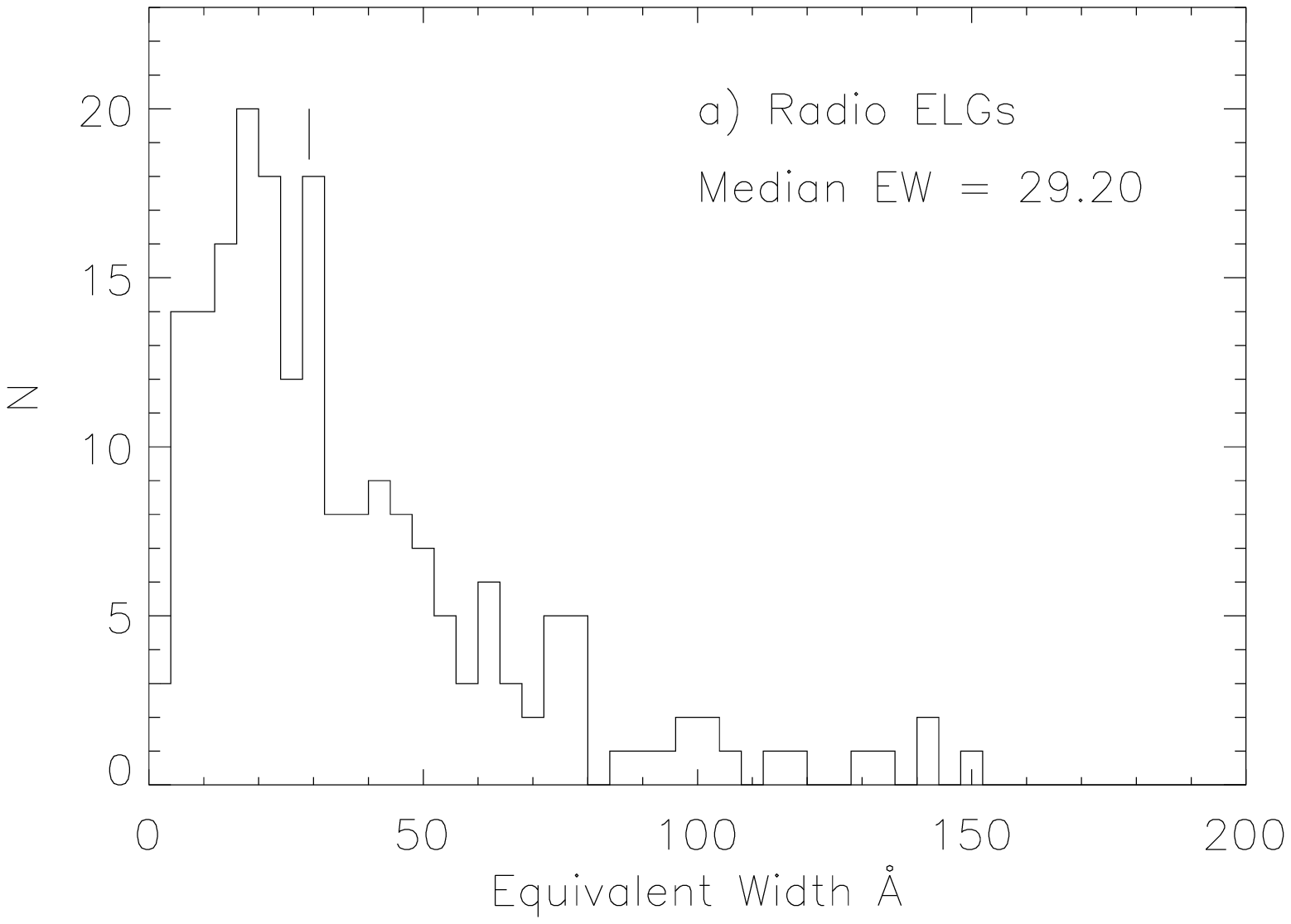}{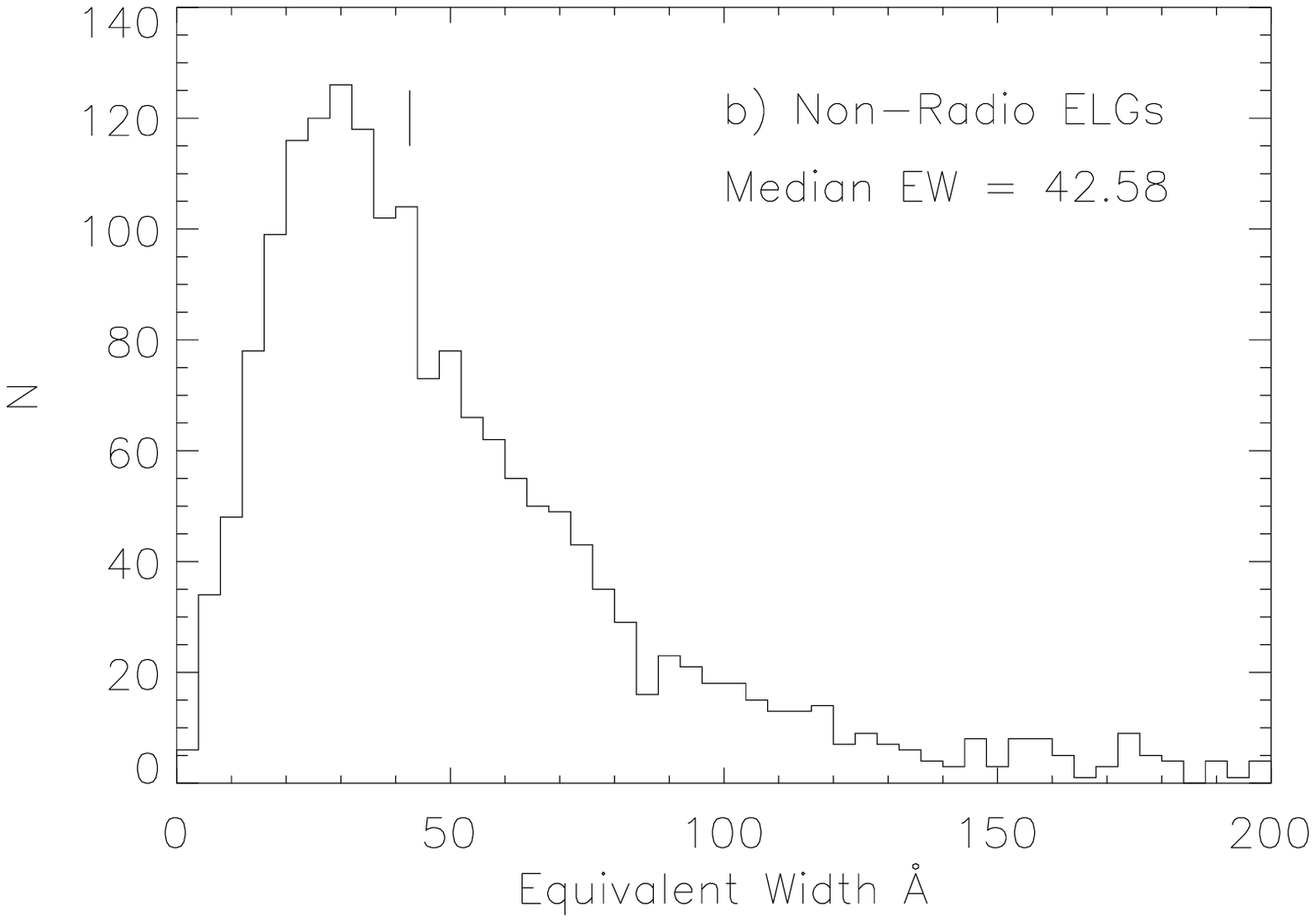}
\figcaption[plots/radio_ew.eps]{Equivalent widths of the H$\alpha$ emission line
  from the objective-prism spectra for (a) radio and (b) non-radio ELGs. \label{fig:ew}}
\end{figure*}

\begin{figure*}[htp]
\epsfxsize=2.5in
\epsscale{1.8}
\plottwo{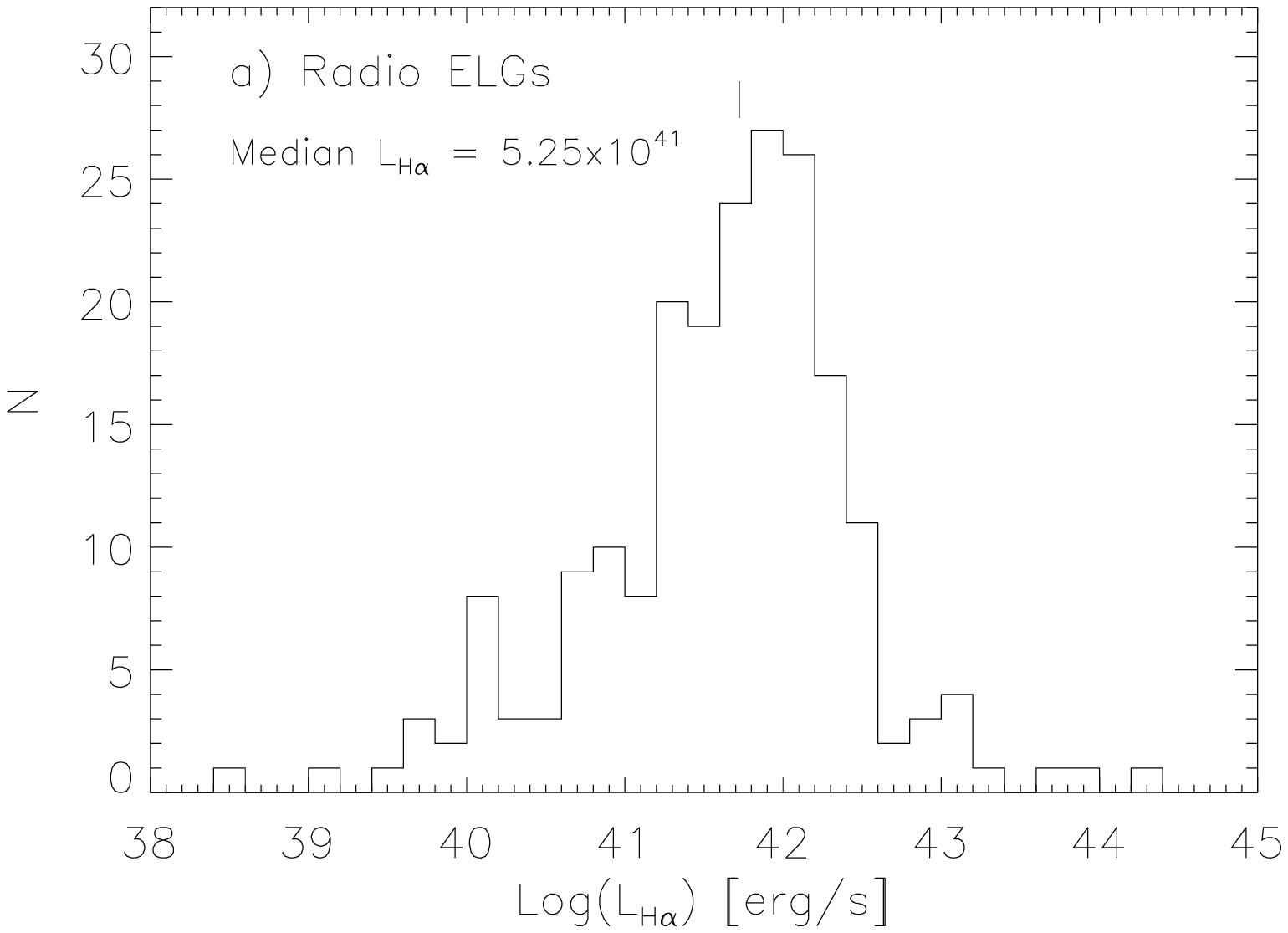}{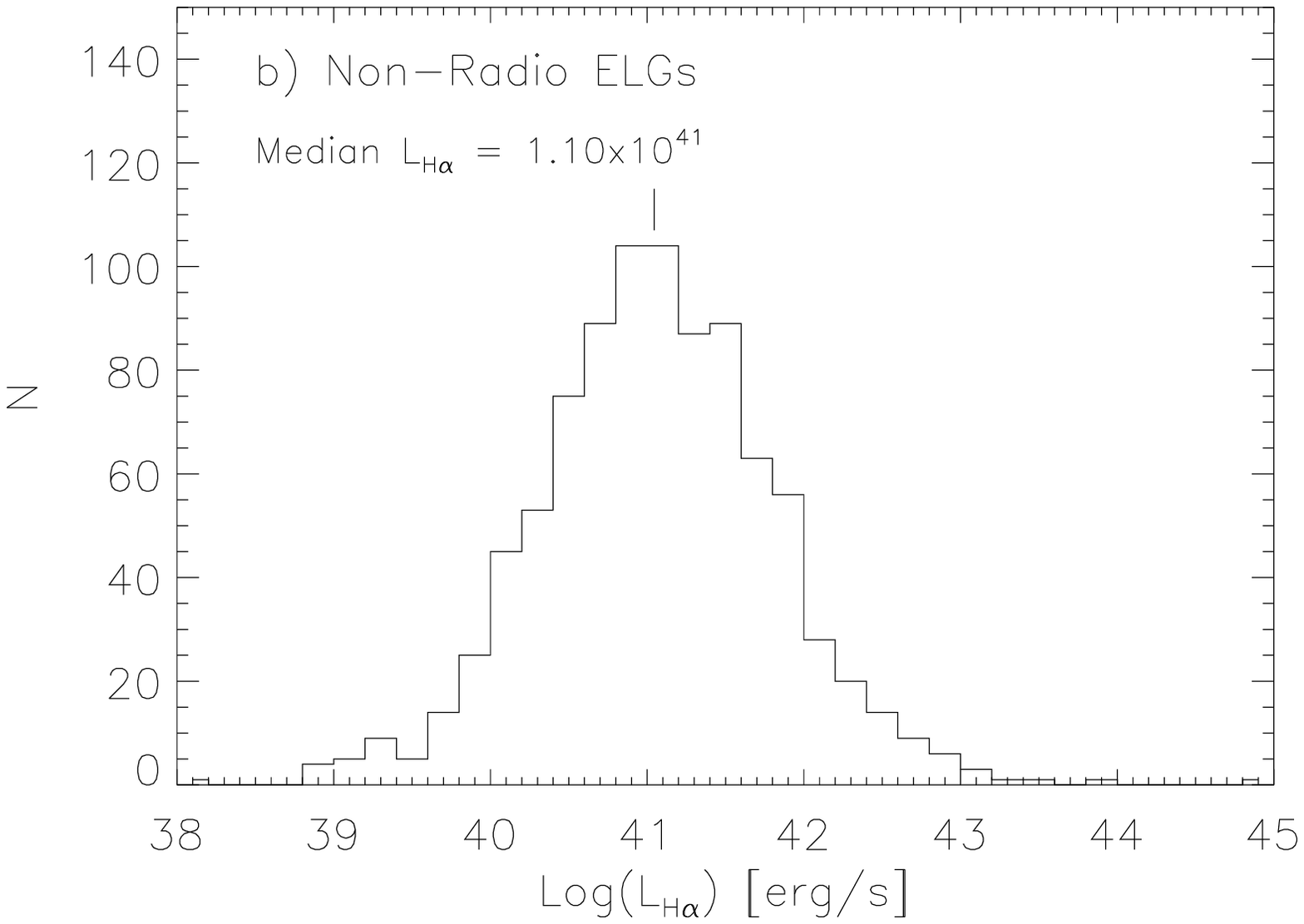}
\figcaption[plots/radio_halum.eps]{(a) Radio and (b) non-radio ELG H$\alpha$
  line luminosities.  \label{fig:halum}}
\end{figure*}

A good measure of the level of star-formation within an ELG is the strength of certain emission 
lines, especially H$\alpha$.  In Figures \ref{fig:ew} and \ref{fig:halum} we show the 
distribution of equivalent widths and H$\alpha$ luminosities, respectively, for both the radio 
and non-radio sub-samples.  The distribution of equivalent widths reveals a selection effect 
inherent in KISS \citep{s00,s01} where the survey is incomplete at low equivalent widths 
(EW $<$ 30 \AA).  This is seen in Figure \ref{fig:ew}b, where the distribution peaks at 
$\sim$30 \AA.  Figure \ref{fig:ew}a reinforces the fact that radio emission favors
high luminosity galaxies (Kellerman \& Owen 1988; Figure 7), which tend to correspond to 
lower equivalent widths.  The median equivalent width of the radio sub-sample is 29.2 \AA, 
compared to 42.7 \AA\ for the non-radio ELGs.  Figure \ref{fig:halum} shows that strong
radio emission tends to be associated with high H$\alpha$ luminosities.  The median 
L$_{H\alpha}$ value for the radio subset is a factor of 5 greater than the non-radio ELGs.  
Discussion of the relationship between L$_{H\alpha}$ and radio power is deferred to a
future paper which presents the IRAS FIR properties of the KISS ELGs and compares 
star-formation rate estimates derived from FIR, radio, and H$\alpha$ luminosities \citep{cg_iras}.

\subsubsection{Line Diagnostic Diagrams}
\begin{figure*}[htp]
\epsfxsize=3.0in
\epsscale{2.0}
\plottwo{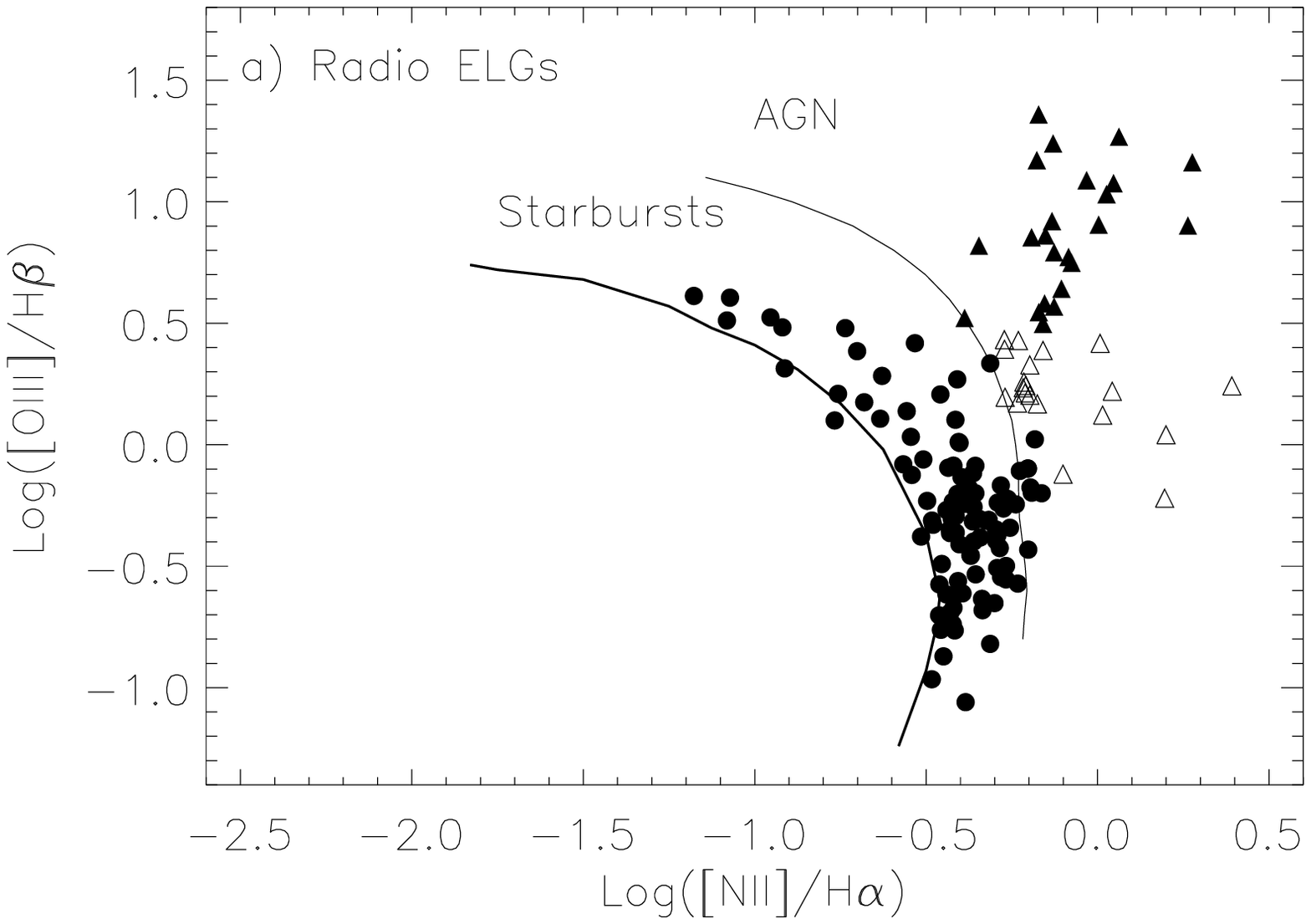}{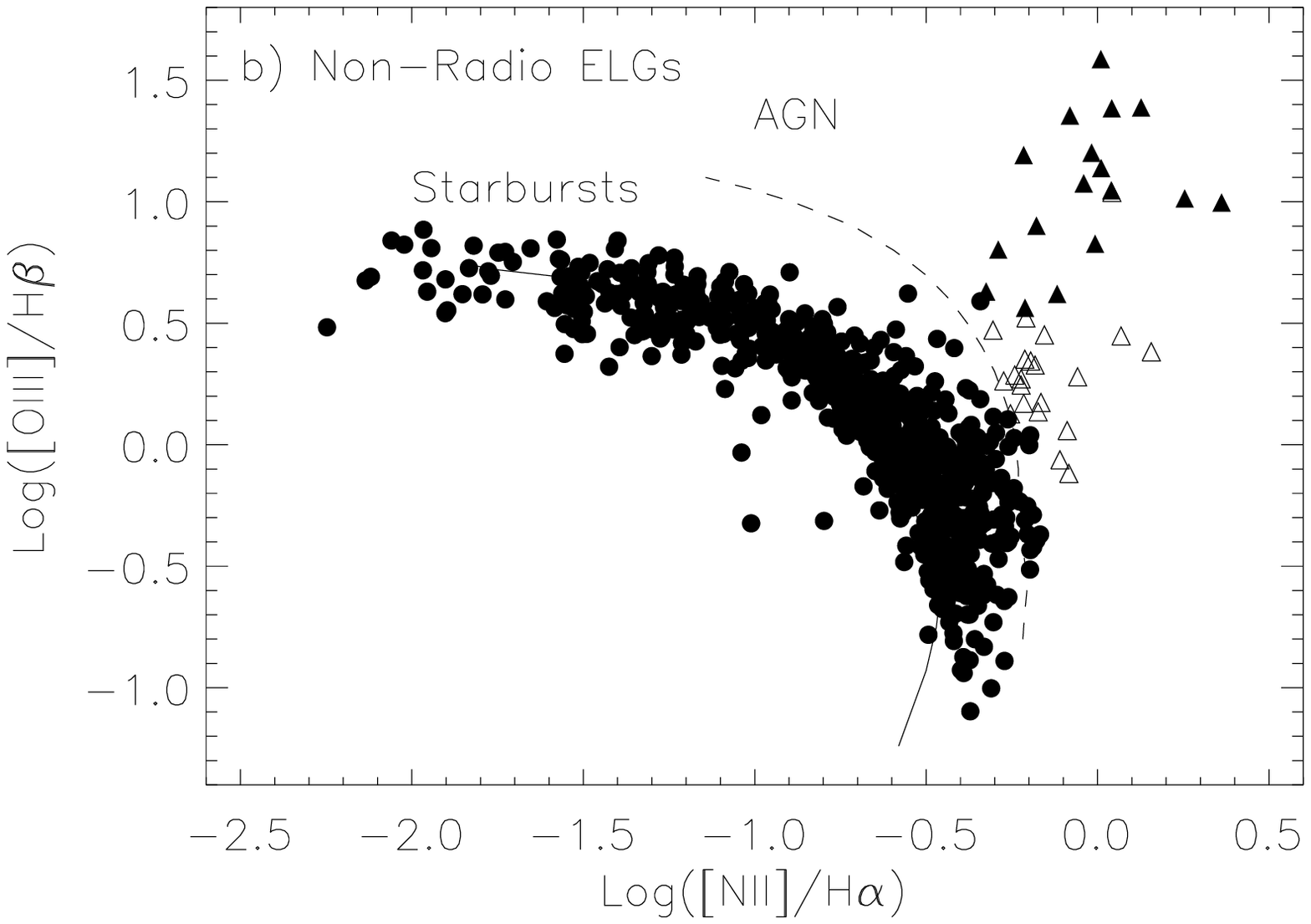}
\figcaption[plots/linediag_all.eps]{Line diagnostic diagrams of the
  (a) radio-detected subset (145 ELGs) and (b) entire KISS sample with
  high quality follow-up spectroscopy (633 ELGs).  ELGs with follow-up
  spectroscopy of insufficient quality to measure all four emission lines
  (H$\alpha$, H$\beta$, [N II], and [O III]) are not displayed.  Symbols
  represent Seyfert 2s (filled triangles), LINERs (open triangles), and
  starbursts (circles). \label{fig:ldd}}
\end{figure*}

The follow-up spectroscopy obtained for the KISS ELGs allows us to generate diagnostic 
diagrams of all galaxies with spectra of sufficient quality to obtain line ratios. 
Figure \ref{fig:ldd} displays a diagnostic diagram for both the radio-detected and full 
KISS samples.  Plotted is the ratio of [OIII]/H$\beta$ {\it vs.} the ratio of [NII]/H$\alpha$.
Different symbols denote different types of active galaxies: Seyfert 2 galaxies (filled
triangles), LINERs (open triangles), and star-forming galaxies (filled circles).  While 
star-forming galaxies typically lie along the HII sequence, AGNs have line ratios of Log([NII]/H$\alpha$) 
$> -0.4$ and lie above and to the right of the star-forming galaxies.  Seyfert 2s are 
located at high values of [OIII]/H$\beta$ ($>$ 3), while the LINERs are located between the 
Seyferts and starburst galaxies \citep{baldwin,veilleux}.


Figure \ref{fig:ldd}b shows all the KISS ELGs with follow-up spectroscopy of sufficient 
quality to measure the necessary line ratios.  While not all the KISS ELGs have follow-up 
spectra, all of the radio subset have follow-up spectra and are represented in 
Figure \ref{fig:ldd}a.  The distribution of Seyferts and LINERs in the two diagrams
is very similar.  This should not be a surprise, since a large fraction of the overall 
KISS AGNs are radio detected.  As mentioned above, 50\% of both the Seyfert 2s and LINERs 
identified in KISS are detected as radio sources.  The
radio-detected starburst galaxies are concentrated toward the high luminosity, high 
metallicity end of the HII sequence, while the full KISS sample populates the sequence
more uniformly.  As mentioned previously, starburst galaxies with detectable radio sources 
are generally limited to higher luminosity galaxies.   These relatively high mass galaxies
have high metallicities, which place them in the lower right section of the HII sequence
(Melbourne \& Salzer 2002).  Compared to the full KISS sample, which spans the full range
of metallicities observed in galaxies, the radio sample includes only a modest number of
intermediate metallicity ELGs (central portion of the HII sequence) and no low abundance
objects (upper left portion of the sequence).

\subsection{Radio Characteristics}

Due to the extremely faint (in the optical) nature of many of their constituents, 
early radio-optical surveys \citep{wind85,benn93} experienced difficulty obtaining
spectroscopy for the majority of their sample.  Thus, these surveys dealt primarily
with the radio flux aspect of the sources.  Recently, radio samples have been 
constructed for which follow-up spectroscopy for most, if not all, galaxies could be 
obtained \citep{pran01,sad02,yrc01}.  Consequently, radio luminosities and activity 
types could be derived for the majority of the galaxies, providing a more complete
understanding of the nature of the galaxies contained in the sample.  All of the 
galaxies in the current study have redshifts and detailed spectral data available.

\begin{figure*}[htp]
\epsfxsize=3.0in
\epsscale{1.0}
\plotone{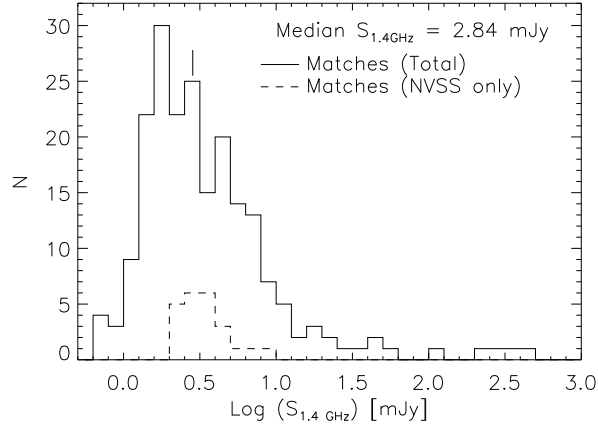}
\figcaption[plots/radflux.eps]{Distribution of the radio flux densities of the 
  combined FIRST and NVSS detections for KISS ELGs.  \label{fig:radflux}}
\end{figure*}

\begin{figure*}[htp]
\epsfxsize=3.0in
\epsscale{1.0}
\plotone{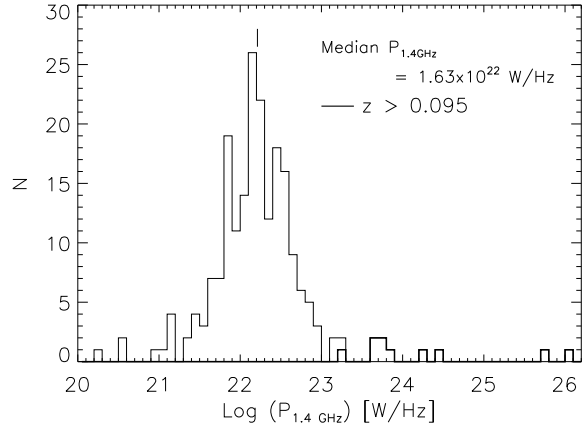}
\figcaption[plots/radlum.eps]{Radio power distribution for the radio
  ELG sample.  All galaxies represented by the heavy histogram have
  redshifts greater than the H$\alpha$ selection limit ($z>0.095$).  \label{fig:radlum}}
\end{figure*}

Figures \ref{fig:radflux} and \ref{fig:radlum} show the distribution of integrated 
radio fluxes and radio luminosities of all 207 radio-KISS ELGs.  The majority of the
sample lies in the 1 - 10 mJy flux range, similar to \citet{kron85} and \citet{wind85}.  
Due to the volume limits imposed by our filter and optical selection method, we do not 
need to go to sub-mJy fluxes \citep{benn93,george99,grupp99,pran01} to obtain a substantial 
sample of nearby, low-luminosity radio-emitting star-forming galaxies.  In Figure 
\ref{fig:radflux}, there are galaxies with integrated fluxes below 1.0 mJy.  While FIRST 
has a stated flux limit of 1.0 mJy, the integrated flux density of FIRST sources can drop
slightly below 1 mJy.   Those galaxies in Figure \ref{fig:radflux} with relatively high 
fluxes ($S_{1.4GHz}\ge100$ mJy) include one very nearby LINER, NGC 4278 (KISSR 29, z=0.0020),
the nearby giant irregular NGC 4449 (KISSR 1307, z=0.00063), and 
two high redshift Seyfert 2s (KISSR 1304, z=0.3481 and KISSR 1561, z=0.3380), which have
radio luminosities typical of classic radio galaxies.  The rarity of high flux sources
in the main KISS volume (z $<$ 0.095) shows how uncommon it is to find extremely bright
sources locally.  This point will be discussed further when we present the radio
luminosity function for the KISS sample in \S5.

\begin{figure*}[htp]
\epsfxsize=3.0in
\epsscale{1.0}
\plotone{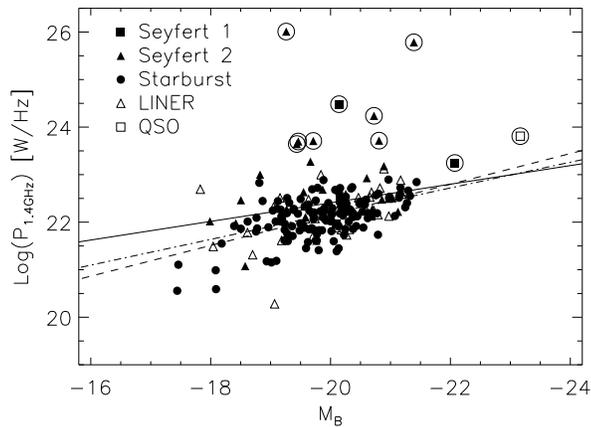}
\figcaption[plots/radlum_mb.eps]{Radio power vs. optical absolute
  magnitude.  The correlation does not appear to be very strong,
  however, a definite trend does exist among the ELG types.  The lines are the 
  fits to the starburst
  (dashed), Seyfert 2 (solid), and LINER (dashed-dotted) populations.
  Circled symbols represent objects detected at $z>0.095$ and are not included
  in the linear least-squares fits.
  \label{fig:radlum_mb}}
\end{figure*}

The radio power distribution (Figure \ref{fig:radlum}) shows a narrow distribution, with 
a median value of $P_{1.4GHz}=1.63\times10^{22}$ W/Hz.  There are 15 galaxies with radio 
powers less than $3\times10^{21}$ W/Hz, which represent the low end of the radio 
luminosity distribution.  All are classified as star-forming galaxies.  At the other extreme,
there are 27 galaxies with radio luminosities greater than $5\times10^{22}$ W/Hz.
However, 10 of these galaxies have high redshifts (z $>$ 0.095) and are not within the 
main survey volume.  These are represented by the heavy histogram in Figure \ref{fig:radlum}.
They will not be included in any further analysis in this paper because they were not 
detected via the H$\alpha$ emission line, but by redshifted [OIII]$\lambda$5007 \AA\ or 
H$\beta$ emission lines.  This high-z population is composed of seven Seyfert 2s, 
two Seyfert 1s, and one QSO.  They are good examples of the radio-loud galaxies typically 
found in standard flux-limited radio surveys.  Excluding these high redshift galaxies, we have 
a relatively well-defined upper limit to our radio luminosity distribution; only three 
galaxies have $P_{1.4GHz}\ge10^{23}$ W/Hz.  Within the KISS volume, we find that there are $no$ 
prototypical ``radio galaxies'' \citep{c92} like those found in standard 
flux-limited surveys.  In such surveys, these extremely luminous galaxies are easy to 
detect out to great distances, but as we show in Figure \ref{fig:radlum}, they are not 
the most common type of radio emitting galaxy in the local universe.  These radio-loud
objects are analogous to O stars in stellar populations, easy to detect due to 
their high intrinsic luminosity, but very rare within a galactic disk.  The radio ELGs 
we observe, on the other hand, are analogous to G, K, and M type stars, far less 
luminous, but much more common in the local volume.  In flux-limited 
surveys the radio-loud galaxies are being oversampled, while the more common radio 
galaxies are under-represented.  This is just an example of the Malmquist effect that is 
seen in flux-limited stellar surveys.  Our result implies that we have greatly reduced 
the sample bias caused by the Malmquist effect and are detecting mostly ``normal'' 
radio-emitting galaxies within the KISS volume.  

Further insight into the physical properties of this sample can be obtained 
through the direct comparison of the  radio power and absolute magnitude for the 
KISS galaxies (Figure \ref{fig:radlum_mb}).  Galaxies are plotted using different
symbols based on their activity types, as indicated by the legend to the figure.
Objects with high redshifts, which tend to be the most luminous objects, are
circled in the plot.  A weak but definite trend of increasing radio power with 
increasing optical luminosity is seen in the low redshift sample.  Linear least-squares 
fits were calculated for each of the various types of ELGs within our sample (excluding 
the Seyfert 1s, which are too few in number to give a reliable fit).   The fits we 
calculate for the Seyfert 2s, LINERs, and starbursts are: 

\begin{equation}
\mbox{Sy2 :}   \log(P_{1.4GHz}) = (-0.20\pm0.11)\cdot M_B + (18.49\pm2.11)
\label{radlum_mb_sy2}
\end{equation}
\begin{equation}
\mbox{SB :}   \log(P_{1.4GHz}) = (-0.32\pm0.04)\cdot M_B + (15.71\pm0.80)
\label{radlum_mb_sb}
\end{equation}
\begin{equation}
\mbox{LINER :}   \log(P_{1.4GHz}) = (-0.27\pm0.10)\cdot M_B + (16.77\pm1.93)
\label{radlum_mb_liner}
\end{equation}

\begin{figure*}[htp]
\epsfxsize=3.0in
\epsscale{1.0}
\plotone{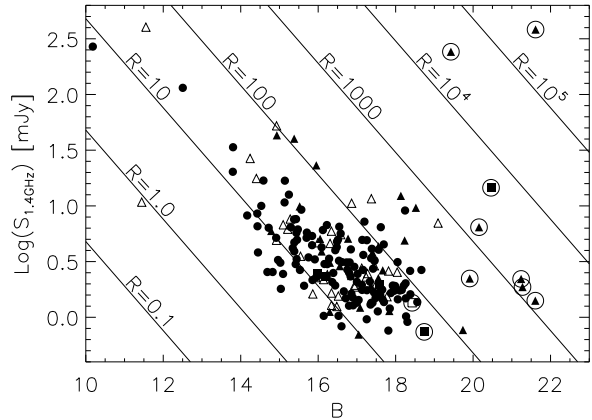}
\figcaption[plots/radopt.eps]{Radio flux density versus B magnitude;
  lines represent constant radio to optical ratio (see
  Eq. \ref{eq:radratio}).  The symbols represent the same ELG types as
  described in Figure \ref{fig:ldd} and include Seyfert 1s (filled
  squares) and one QSO (open square). \label{fig:radopt}}
\end{figure*}

The galaxies appear to follow a linear relationship, though not as tight as the 
well-documented radio power-FIR relationship \citep{c92,yrc01}.  There is also a 
modest trend toward more luminous galaxies in the optical being more luminous in 
the radio wavelengths.   The fits of the starburst and LINER components of the 
sample are almost identical, while the Seyfert 2s tend toward slightly higher radio 
powers.  All fits are the same within the formal uncertainties, due to the scatter 
of each type of galaxy about the mean relation.  These fits do not include the high 
$z$ Seyferts, represented by the circled symbols.  A possible cause for the very 
similar fits between the starburst and LINER populations is that the LINERs we detect
as radio sources may be driven more by star-forming activity than AGN activity.  
However, this is purely speculative.  Moreover, given the small difference between the 
Seyfert 2 and LINER fits, the alternative hypothesis that the LINERs share the same 
activity source as the Seyferts remains plausible.  It is worth noting that few 
surveys of this type have detected such a large LINER population.   Multiwavelength 
observations of the current sample of KISS LINERs may be useful in unraveling the 
enigmatic nature of these objects.

An unexpected result from our survey is that of the comparative distribution of ELGs
by type.  One might initially expect AGNs to dominate at higher radio power or at
higher optical luminosities and starbursts to be concentrated near the fainter end, 
but Figure 13 shows this is not the case.  The distribution of each type of ELG appears 
quite similar.   This means that while Seyferts typically are more luminous when 
compared to starburst ELGs, this is not necessarily the case in the radio wavelengths.  
The same is true for LINERs.  Another interesting result is that at lower radio and 
optical luminosities, the number of galaxies dwindles, and then abruptly drops to zero.  This 
is not due to a lack of galaxies at those optical luminosities within the KISS volume.  
On the contrary, there are a large number of lower luminosity and dwarf galaxies that 
have been classified as actively star forming, but do not appear to have detectable radio 
emission.  This could suggest that radio emission at sufficient levels to be detected
by FIRST or NVSS is only possible beyond a certain luminosity threshold.   Since the
1.4 GHz emission from star-forming galaxies is dominated by synchrotron radiation
which requires a galactic-scale magnetic field, it is possible that these low-luminosity 
galaxies with starbursts and star-forming regions do not emit in the radio due to a lack 
of a sufficiently strong magnetic field.  Therefore, while ELG type may not constrain the 
radio luminosity of a galaxy, the mass of the host galaxy may.

Another analysis tool we employ is the comparison of
radio flux densities against the B magnitudes.   Such a plot is shown in
Figure 14 for the full KISS sample.  Superimposed are lines representing 
constant values for the radio-to-optical ratios, $R$, defined by \citet{c80} as
\begin{equation}
R = S_{1.4} \cdot 10^{0.4(B-14.2)}
\label{eq:radratio}
\end{equation}
where $B$ is the apparent magnitude.  The equation is offset in the
exponent from the original equation derived by Condon to better
compare our $B$ filter magnitudes to the red and near-IR filters
employed by \citet{pran01} and \citet{yrc01}.  Our results show a
strong clustering of galaxies in the range of $1.0 < S_{1.4} < 6.3$
mJy and $10 < R < 100$.  Again, we see that the various types of ELGs 
overlap each other in the diagram.  To be sure, the objects with R 
$\ge$ 100 have tendency to be AGNs (10 of 17 low redshift objects and 
8 of 8 high redshift (circled) objects).  However, between R = 10 and
100, Seyferts, LINERs, and starbursts are coincident in the diagram,
showing that despite the differences in the origin of the radio
emission, the various ELG types exhibit radio emission of similar
strength.  Below R = 10, the majority of the objects are starbursts
(16 of 22), although a few Seyferts and LINERs exist.

As previously noted, the flux limit of FIRST  is not deep enough to directly 
explore the sub-mJy regime that \citet{george99}, \citet{grupp99}, and 
\citet{pran01} cover, but with the relatively high density of local 
radio-detected galaxies, we believe we are detecting the same population 
of lower radio luminosity galaxies found at these sub-mJy fluxes.  For example,
consider the spectroscopic follow-up data given by
\citet{grupp99}.  They attempt to ascertain the nature of the sub-mJy
population of the Marano Field from a sample of 68 faint radio sources
($S > 0.2$ mJy).  They were successful in obtaining redshifts and
activity classes for 34 of these objects.  If we examine only their
late-type spiral class, which equates to our starburst or star-forming
class and accounts for 11 of the 34 objects, the redshift range is $
0.154 \le z \le 0.368$.  The flux range of these galaxies is $ 0.28
\le S_{1.4} \le 0.60$ mJy, with one brighter galaxy at $S_{1.4}=1.71$ mJy.
Considering that all these galaxies lie beyond the KISS redshift
range, if we were to scale their distances to the median redshift of
our sample, the observed flux would be well above our detection limit,
and brighter than many galaxies in our sample, regardless of type.
One late-type galaxy with a radio flux of $S_{1.4} = 0.28$ mJy, an optical
magnitude of $V=18.48$, and a redshift of $z=0.255$, is 2.4 times as far 
as the KISS volume limit of $z=0.095$.  If the object is scaled to this 
limiting redshift it would have an observed flux of 1.6 mJy, well within 
the detection range of FIRST or NVSS and easily detectable as an ELG by 
KISS.  Thus, if this object is a typical sub-mJy detection for flux-limited 
surveys, we are apparently detecting the same population of galaxies.

As another example, we detect the starbursting and star-forming late-type galaxy 
population that dominated at the sub-mJy flux level detected by \citet{pran01} 
within the same $R$ range.  The majority of AGNs detected by \citet{pran01}
are also within $10 < R < 100$, as are the majority of radio-KISS
sample AGNs.  When we ignore the high redshift galaxies, indicated by
the circled symbols, we find that no radio-detected KISS ELG has an
$R\ge1000$ and only a handful have $R\ge100$.  This is the regime that
is found to be dominated by early-type galaxies, which are detected
in large numbers by \citet{pran01}.  However, within our volume of $z\le0.095$,
they detect only 2 early-type galaxies.  Thus, the domination of early-type 
galaxies in the mJy regime may be due to the Malmquist effect, in which these 
luminous galaxies are detected \emph{en masse} at higher redshift but may not 
be at all common in the local volume.  Amplifying this effect is the observed
cosmic evolution of radio galaxies, whose volume density increases with 
increasing redshift \citep{c89}.  During the process of matching the radio 
catalogs with optical KISS counterparts, many 
strong radio sources were found within the KISS fields that were not associated 
with an ELG but did have optical counterparts, most of them faint objects.  
Either these objects possess emission lines that are fainter than the optical 
detection limits of KISS, they lie at redshifts beyond 0.095, or they are early-type
galaxies with strong radio sources but weak or absent emission lines.  
Work has begun in obtaining spectra for the brighter radio objects within 
the KISS fields.  Preliminary results indicate that many are indeed early-type 
and most likely elliptical galaxies, although most lie at redshifts beyond 
0.095.  Once these objects have all been observed spectroscopically we will
be able to account for this missing population in the radio luminosity 
function within the KISS volume, leading to a truly complete volume-limited
luminosity function (see below).

\section{The Local Radio Luminosity Function at 1.4GHz}

A useful way to visualize the make-up of any extragalactic sample of objects
is through the construction of that sample's luminosity function (LF).  The
KISS ELG sample is particularly well suited for this task, since its completeness
limits and selection function are both well understood and readily quantified
using the survey data.  Due to the digital nature of the survey,
KISS is superior in this regard when compared to previous objective-prism 
surveys for active galaxies carried out using photographic plates.

The calculation of the radio luminosity function (RLF) for the KISS ELGs is
dependent on the careful definition of the limiting volumes of each galaxy 
in our sample.  Each galaxy has associated with it three different selection 
limits which reflect  the ability of both the KISS and radio surveys to detect 
it.  They are the limiting emission-line flux level of KISS, the 
filter-imposed redshift limit of KISS, and the radio flux limit of either FIRST 
or NVSS (whichever is appropriate).  The relevant volume that defines the contribution 
of each source to the RLF is determined by a multivariate selection function, defined 
by these three observational limits.  Specifically, the three limiting volumes are
(1) the limiting optical volume, $V_{opt}$, which is dependent on the completeness 
limit of the sample in the optical, the absolute brightness of the galaxy determined 
from the emission-line intensity, and the redshift
of the galaxy; (2) the volume limit of KISS, $V_{KISS}$, which is imposed by the 
H$\alpha$ filter bandpass; and (3) the limiting radio volume, $V_{rad}$, which is
set by the radio flux limit of the survey from which the flux was measured (either
FIRST or NVSS) and the derived radio power of the source.  The volume set by each 
of these limits is computed from the maximum distance that a galaxy could have and
still be detected by each technique.  The final volume used for each galaxy in the RLF
computation is the smallest of the three volumes described above.  These volumes 
are then used to calculate the space density within each luminosity bin of the RLF.

For most luminosity functions that are derived from a flux-limited sample,
correcting for the lack of completeness in the sample is necessary in order
to accurately portray the volume densities of galaxies in the sample.  This
is particularly true for low luminosity galaxy samples that usually
suffer from incompleteness.  This incompleteness is minimized in the
KISS RLF, due to the nature of the survey.  The
majority of the radio-detected sources have V$_{KISS}$ as their minimum
volume.  However, the limiting redshift set by the KISS filter is not a 
hard limit.  As is seen in Figure \ref{fig:z}, the number of galaxies in the 
survey begins to drop at $z>0.085$.  This is due to the fact that galaxies 
with weaker emission lines start to drop out of the sample as their lines 
redshift out of the bandpass of the objective-prism data (Salzer et al. 2000).  
To ensure the most complete sample possible for the calculation of the RLF, we 
use only objects with $z\le0.085$, which totals 184 galaxies.  This eliminates 
the 10 high $z$ galaxies from the sample (which we would have excluded anyway), 
as well as 13 galaxies with redshifts between $z=0.085$ and 0.095.  All of these 
latter galaxies have radio luminosities very near the median value, meaning that 
we are not biasing the resulting RLF by excluding these sources.  In addition, we
have removed two ELGs with very small redshifts ($z<0.0025$) since their small
limiting volumes cause them to make inappropriately large contributions to the
computed volume densities in their respective luminosity bins.  Hence, the total
sample used for the RLF derivation is 182 KISS ELGs.

\begin{figure*}[htp]
\epsfxsize=4.0in
\epsscale{1.5}
\plotone{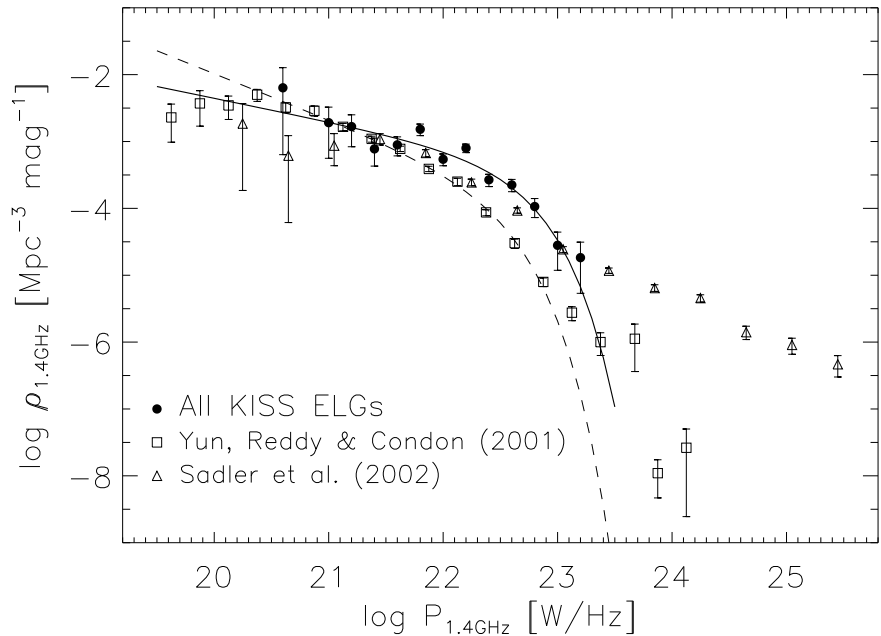}
\figcaption[plots/rlfcombo_all.eps]{Radio luminosity function (RLF)
  derived for the KISS ELGs in the local volume of $z\le0.085$.  The 
  filled circles and solid line show our own data and the associated
  Schechter function fit.  The squares and dashed line show \citet{yrc01}
  data and associated Schechter fit for low luminosity galaxies.  The
  triangles represent the RLF from \citet{sad02}.  \label{fig:RLF}}
\end{figure*}

We adopt the cosmology of $H_0 = 75$ km s$^{-1}$ Mpc$^{-3}$ and $\Omega_M = 1.0$, 
$\Omega_{\Lambda} = 0$, and $\Omega_k = 0$ for a flat, $\Lambda = 0$, universe throughout 
all our calculations.  These choices allow us to directly compare our RLFs with RLFs from 
previous papers.  The space density 
at each luminosity bin is calculated via the 1/$V_{max}$ method \citep{schmidt68}.
We compute $V_{KISS}$, $V_{opt}$, and $V_{rad}$ for each source and substitute the 
smallest of these quantities for $V_{max}$ to obtain the final limiting volume for 
a particular galaxy.  Using our adopted KISS redshift limit of 0.085, the limiting 
KISS volume (V$_{KISS}$) for the luminosity function is computed to be $2.1 \times 
10^5$ Mpc$^3$.  This is the maximum volume for any galaxy within our survey.  The 
calculation of $V_{opt}$ is based on the observed emission-line flux of each source, 
as measured from the objective-prism spectra.  Using the line fluxes for the full
KISS sample, we have derived a completeness limit for the survey, which is used to
evaluate the maximum distance out to which each source can be detected based on its 
line strength.  This distance is then used to compute $V_{opt}$.  Full details, including
the derivation of the limiting line flux of the KISS sample, is given in \citet{kisscomplete}.  
The radio visibility function is calculated by solving for the distance to the galaxy 
in terms of the radio luminosity and calculating the radio visibility for each galaxy via 
\begin{equation}
V_{rad} = 1.011\times10^{-25} \left(\frac{P_{1.4}}{S_{lim}}\right)^{3/2}
\end{equation}
where V$_{rad}$ has units of Mpc$^3$, P$_{1.4}$ is in units of W/Hz, and 
$S_{lim}$ is the radio flux limit for the galaxy.  If the galaxy is detected by 
both FIRST and NVSS or only by FIRST, $S_{lim} = 1.0$ mJy.  If detected only by 
NVSS, $S_{lim} = 2.5$ mJy.   

Once all three selection limits are taken into account, each of the 182 KISS ELGs
included in the RLF calculation is assigned into one of three subsamples.  The
KISS volume-limited subsample, where the volume is set by V$_{KISS}$, includes
98 objects (54\% of the total).  The radio-flux-limited subsample accounts for an 
additional 66 ELGs (36\%).  The smallest subset is the optical emission-line-flux-limited
subsample, which comprises only 18 objects, or 10\% of the overall sample.

Since we employ the 1/$V_{max}$ estimator for computing our luminosity functions, we
are susceptible to variations in the large-scale structure (LSS) within our survey
volumes.  In particular, the second survey strip includes a section of the Bo\"otes void.
One might assume that the presence of this low-density region will reduce the 
overall amplitude of our RLF.  However, we stress that the void covers less than 4\% of
the volume of the 2nd survey strip \citep{kiss43}, and hence less than 2\% of the overall 
volume used for computing the current RLF.  Furthermore, the depth of the KISS survey
ensures that variations in the LSS are effectively averaged over since the survey
samples the galaxian distribution over at least a few coherence lengths.  Therefore, we do 
not believe that LSS variations are dramatically affecting our RLF.  This belief is
supported by our computation of an optical LF for the first survey strip \citep{kisscomplete}
using both the 1/$V_{max}$ method and the inhomogeneity-independent Lynden-Bell ``C" method
\citep{LB71,dLGH89}.  The LFs computed via the two independent methods were indistinguishable
in terms of their shapes, suggesting that LSS variations are not causing significant
problems when only the 1/$V_{max}$ method is used.

\begin{figure*}[htp]
\epsfxsize=4.0in
\epsscale{1.5}
\plotone{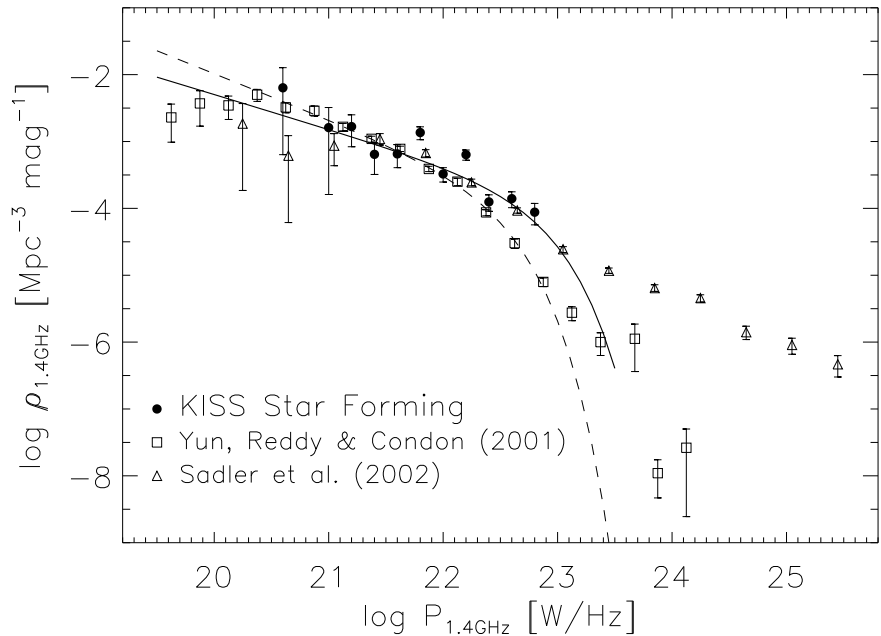}
\figcaption[plots/rlfcombo_sb.eps]{KISS RLF derived for the starburst
  galaxies only (filled circles).  The additional RLFs and Schechter fits
  plotted are the same data sets as described in Figure \ref{fig:RLF}. 
  \label{fig:RLF_SB}}
\end{figure*}

Figure \ref{fig:RLF} shows the derived KISS RLF, plotted as filled circles.
For comparison, we also plot the RLFs  derived by \citet{yrc01} (henceforth YRC,
open squares) and \citet{sad02} (henceforth S02, open triangles).  The YRC sample
is derived from a correlation of the NVSS with a catalog of IRAS-detected galaxies,
and includes 1809 galaxies with redshifts as high as 0.16.  Hence, it is selected via a
combination of the IRAS and NVSS selection functions and flux limits. The S02 sample 
represents those galaxies from a subset of the 2dF redshift survey detected by NVSS, 
and consists of 757 galaxies with redshifts out to 0.44.  The selection function for
this sample is thus a combination of the NVSS and 2dF limits (see S02).   The S02
RLF has been corrected to $H_0 = 75$ km s$^{-1}$ Mpc$^{-3}$.  The solid line in
Figure \ref{fig:RLF} represents the Schechter (1976) function fit to the KISS RLF, 
while the dashed line is the Schechter function fit to the lower luminosity galaxies 
only (P$_{1.4}$ $<$ 10$^{23.2}$ W/Hz) of YRC.  The Schechter function has the form
\begin{equation}
\rho(L)dL=\rho^* \left(\frac{L}{L^*}\right)^{-\alpha}
exp\left(-\frac{L}{L^*}\right)d\left(\frac{L}{L^*}\right).
\end{equation}
The parameters $\rho^*$ and $L^*$ are the characteristic density and luminosity 
of the population and $\alpha$ describes the faint-end power-law slope for 
$L << L^*$.  Figure \ref{fig:RLF_SB} shows the RLF for just the KISS star-forming
galaxies, along with the same RLFs from YRC and S02 for comparison.  The volume
densities from both KISS RLFs are listed in Table \ref{tab:RLF}, along with the
corresponding numbers for the AGNs (not plotted separately).  The formal errors
listed in the table and illustrated in the two RLF plots are based on $\sqrt{N}$
uncertainties in each luminosity bin.  Table \ref{tab:lf_sol} 
lists the Schechter parameters for the full radio sample and the starburst-only RLFs,
along with the values obtained from the fit to the YRC data.  We do not attempt to
fit the S02 RLF, as it does not appear to be well represented by the shape of the
Schechter function.



The full KISS galaxy RLF (Figure \ref{fig:RLF}) agrees well with the RLF
for the IRAS-NVSS sample of YRC.  Below P$_{1.4}$ = 10$^{22}$ W/Hz the two RLFs
lie on top of each other.  Above P$_{1.4}$ = 10$^{22}$ W/Hz the KISS RLF exceeds
that of YRC by a modest amount out to the last point plotted at P$_{1.4}$ = 
10$^{23.2}$ W/Hz.  Most of the excess between the KISS and YRC RLFs above P$_{1.4}$ 
= 10$^{22}$ W/Hz is probably due to sample differences.  The IRAS-based sample of 
YRC is made up predominantly of star-forming galaxies ($\sim$99\% according to
YRC), while the KISS radio-detected sample includes a large contribution from AGNs.  
Fully 33\% of the redshift-restricted sample of 182 galaxies used to derive the
KISS RLF are AGNs.  When the RLF is computed for the star-forming galaxies alone
(Figure \ref{fig:RLF_SB}), the YRC and KISS RLFs agree even more closely.  In
both cases, the YRC RLF reaches to higher radio luminosities than does the KISS
sample.  This is most likely due to the fact that the KISS sample is truncated
in redshift.  This redshift limit greatly restricts the distance and hence the volume
that KISS is probing, meaning that KISS is not sensitive to the relatively rare
radio-luminous objects.  As seen in Figure \ref{fig:radlum}, there are in fact
KISS galaxies with P$_{1.4}$ $>$ 10$^{23.5}$ W/Hz.  However, these are all higher
redshift objects that are not part of the quasi-volume-limited sample being used
to construct the RLF.  There are no radio loud galaxies within the KISS
volume that have emission-line strengths high enough to be selected by KISS.
The YRC sample probes roughly twice as deep in redshift and detects some rare 
radio luminous starburst galaxies at P$_{1.4}$ $>$ 10$^{23.5}$ W/Hz.

The similarities and differences between the KISS and YRC RLFs are echoed in the 
Schechter function parameters listed in Table \ref{tab:lf_sol}.  For the full KISS
RLF, the Schechter fit exhibits both a shallower faint-end slope and a somewhat higher
value for L* than the YRC data.  The latter is almost certainly due to the significant
contribution from the AGNs to the KISS RLF.  When the AGNs are removed and only
the KISS star-forming galaxies are considered, the faint-end slope $\alpha$ more
nearly agrees with the value derived from the YRC sample; they are consistent with
each other within the uncertainties.  Note that the value of L* for the KISS 
star-forming RLF is poorly constrained by the data and must be treated with caution
since the formal value for L* is located at the highest-luminosity bin in the
RLF.  In general the uncertainties in all of the fitting parameters are higher
for the KISS RLFs than the YRC RLF, due to the smaller number of objects present
in the KISS radio sample.

The situation is different for the comparison between the KISS and S02 
RLFs.  The latter exhibits a shallower faint-end slope than either the KISS or YRC 
RLF, and lies significantly below the other two at low radio powers.  However, the 
S02 and KISS functions agree quite well at intermediate luminosities (between P$_{1.4}$ 
= 10$^{21.4}$ W/Hz and P$_{1.4}$ = 10$^{23.2}$ W/Hz).  Then, at radio powers above 
10$^{23.2}$ W/Hz, the S02 RLF greatly exceeds both the KISS and YRC functions.  The 
S02 sample is derived from the deep 2dF redshift survey, and includes objects out to 
z = 0.438.  The majority of the S02 sample are AGNs, particularly those with redshifts 
beyond z = 0.15 where $\sim$90\% of their radio detections correspond to AGNs.  Because 
of the large volumes covered by the S02 sample, they detect many more of the rare 
radio-loud galaxies that are largely excluded by the redshift limit of KISS.  Thus, 
while S02 appears to be under-sampling the radio population at low powers, it is quite
sensitive to the very powerful radio galaxies.  The large differences between the YRC and S02 RLFs 
at P$_{1.4}$ $>$ 10$^{23}$ W/Hz emphasizes the fact that essentially all radio-loud 
objects in the universe are AGNs.  Note that when the S02 sample is broken down into
AGNs and star-forming subsamples, the RLF for the star-forming galaxies does agree
well with the KISS starburst RLF at low and intermediate luminosities (see Figure 15
of S02).

One might be concerned that the KISS sample may be missing substantial numbers
of radio sources that lie within the survey volume but which have either weak
or no emission lines.  After all, a large number of radio-loud galaxies exhibit
spectra characteristic of elliptical galaxies.  However, since the KISS RLF
agrees well with both the YRC and S02 RLFs in the radio power range covered by
KISS, it would appear that this is not a substantial problem below P$_{1.4}$ = 
10$^{23.2}$ W/Hz.  At higher radio luminosities, the volume densities found by
S02 are below 10$^{-5}$ Mpc$^{-3}$ per luminosity bin.  Since the volume limit
of KISS is $\sim$2 $\times$ 10$^5$ Mpc$^3$, the number of radio sources that
might be missed by KISS is $\sim$2 or less in each of the luminosity bins of
Figure \ref{fig:RLF} ($\sim$5 -- 6 galaxies total).  Hence, the KISS RLF is
essentially undetermined for P$_{1.4}$ $>$ 10$^{23.2}$ W/Hz.   It would clearly
be of interest to know more precisely which radio sources in the KISS volume are
missed by the optical objective-prism survey, and what their properties are.
For example, Miller \& Owen (2002) found a large population of cluster galaxies 
that seem to host dust-obscured starbursts yet whose spectra are devoid of emission 
lines.  Knowing how significant such a population of optically-obscured 
star-forming galaxies might be would have important implications for our
understanding of galaxy evolution and the local star-formation rate density.
As mentioned in the previous section, we are in the process of obtaining spectra 
of the optically bright galaxies with radio detections that lie in the KISS regions.  
To date, only two objects with radio detections have been identified within the KISS
volume, consistent with the estimates above.  We will continue to pursue this issue
with additional spectroscopy in the future.


\section{Summary and Conclusions}

We have taken advantage of the existence of three unique wide-field
surveys to carry out a multi-wavelength study of a deep sample of active galaxies.
Using the optically selected emission-line galaxies from the first two H$\alpha$-selected
lists of the KISS project, we have correlated the positions of 2157 ELGs with both
the FIRST and NVSS 1.4 GHz radio surveys.  Among the goals of this study are to determine
the incidence of detectable radio emission from star-forming galaxies and AGN, and
to develop a picture of the characteristics of radio galaxies in the local universe
(z $<$ 0.1).  While most studies based on radio surveys have focused on the
optical characteristics of flux-limited samples of radio sources, we instead desire
to probe the radio characteristics of an optically-selected, quasi-volume-limited sample 
of ELGs.

Our positional matching exercise yielded a total of 207 radio detections (9.6\%
of the full KISS sample).  Of these, 184 were detected in FIRST and 147 in NVSS.
We used a variety of visual checks to evaluate the reality of all possible matches with
radio-optical separations of less than 30 arcsec.  Therefore, we feel that our final
catalog of radio-detected KISS ELGs should be extremely reliable.  The median
positional difference between the FIRST and KISS galaxies is 0.75 arcsec, with only 5
matches being more that 2 arcsec apart.  For the NVSS-KISS detections, the median
separation is 3.4 arcsec (reflecting the lower spatial resolution of NVSS), with all but 2
of the radio sources located within 15 arcsec of the optical target.

An important aspect of our study is that all of our 207 radio-detected ELGs
possess a follow-up spectrum in the KISS spectral database.  These spectra
provide confirmation that the KISS ELG candidates are in fact a {\it bona fide} 
emission-line source, plus yield accurate redshifts and line ratios.  The latter are
employed as diagnostics of the type of activity present in each galaxy.

To gain insights into the nature of the KISS radio galaxies,
we first compare the optical properties of the radio-detected ELGs with the large
number of radio-quiet KISS galaxies.  The redshift distributions of the radio and
non-radio populations are indistinguishable, indicating that the radio subsample
is not biased with respect to distance.  However, the radio-detected sources are
on average significantly brighter (median B magnitude of 16.84 compared to 18.21) 
and more luminous (median B-band absolute magnitude of $-$19.89 compared to 
$-$18.87) than the non-radio KISS ELGs.  Few radio-detected objects have B
magnitudes fainter than 18.5, and all of these are higher redshift sources.  The 
faintest object in the radio sample has B = 21.61.  The radio-detected galaxies also 
tend to have higher H$\alpha$ luminosities but lower H$\alpha$ equivalent widths,
compared to the non-radio objects.

The availability of follow-up spectra allow us to evaluate the activity type of each
radio-detected KISS galaxy.  Using standard emission-line diagnostics, we determined 
that 132 of 207 galaxies (63.8\%) are starburst/star-forming ELGs, 37 of 207 are LINERs
(17.9\%), 32 of 207 are Seyfert 2s (15.5\%), 5 of 207 are Seyfert 1s (2.4\%), and 1
object is a quasar (0.5\%).  In total, 75 of 207 are AGNs of some type (36.2\%), which
is much higher than the overall proportion of AGNs among those KISS objects with
follow-up spectra (14.7\%).  However, the proportion of the radio sample that are
AGNs is {\it low} when compared to radio-selected samples of objects (see below).

The star-forming galaxies in the radio subsample represent a biased population when
compared to the overall population of KISS star-forming galaxies.  In the line diagnostic
diagram they tend to lie in the lower excitation portion of the star-forming sequence.  
These are galaxies with higher luminosities and higher metallicities compared to the 
overall KISS star-forming galaxy population.  The many intermediate and lower luminosity 
KISS star-forming galaxies are completely absent from the radio-detected population.   
We interpret this as being due to, at least in part, the relative weakness of galaxian-scale 
magnetic fields in dwarf galaxies.   

The median radio flux of the combined FIRST and NVSS radio detections is 2.84 mJy.  
Four sources have radio fluxes in excess of 100 mJy: KISSR 29 (= NGC 4278), which
is one of the brightest objects in the KISS catalogs, and is a nearby (z = 0.0020) 
elliptical galaxy with a LINER spectrum, KISSR 1307 (= NGC 4449), the well studied
nearby (z = 0.00063) Magellanic irregular galaxy, and KISSR 1304 and KISSR 1561, both high
redshift Seyfert 2 galaxies (z = 0.3481 and 0.3380, respectively) that are also the
galaxies with the highest radio powers.  The median radio power for the full sample
is P$_{1.4 GHz}$ = 1.63 $\times$ 10$^{22}$ W/Hz, with the vast majority of the KISS objects
having powers in the range 10$^{21.5}$ to 10$^{22.7}$ W/Hz.  Only nine KISS ELGs have
radio powers in excess of 10$^{23.5}$, and all are high redshift objects.

An interesting aspect of the properties of the KISS sample is that the ELGs of different 
activity types strongly overlap in terms of their radio characteristics.   The starbursting
ELGs have essentially the same levels of radio power and radio-optical ratios  as
the LINERs and Seyferts.  In other words, one cannot use the radio luminosity or some
other radio parameter to distinguish between the various activity types.  Based on previous
studies of radio-flux-selected samples, one might have expected that the AGNs would
have, on average, higher radio powers.  However, in our local, quasi-volume-limited
sample, the starburst galaxies are contributing just as much to the total radio power as
are the AGNs.  Within our volume (z $<$ 0.095) there are no extremely powerful
radio galaxies.

A radio luminosity function (RLF) was constructed for the full KISS ELG sample,
as well as for the star-forming and AGN subsamples.   A total of 182 objects with
z $<$ 0.085 were used for the computation of the RLF, including 122 star-forming 
galaxies (67.0\%) and 60 AGNs (33.0\%).   Our RLFs agree well with those computed
by \citet{yrc01} for a sample of IRAS galaxies also detected by NVSS, and by
\citet{sad02} who matched the deep 2dF redshift survey with the NVSS.  The \citet{yrc01} 
sample complements ours, in that the galaxies are selected as having FIR IRAS
emission, meaning that they are typically later-type galaxies with a substantial ISM.
These authors indicate that AGNs make up only a small percentage of the IRAF-NVSS
sample.  Therefore it is no surprise that the KISS star-forming galaxy RLF agrees
extremely well with that of \citet{yrc01}.  The \citet{sad02} sample contains a large
population of strong radio sources at modest redshifts, but still agrees
quite well with the KISS RLF in the intermediate radio power range (P$_{1.4 GHz}$ 
between 10$^{21.4}$ and 10$^{23.2}$ W/Hz).

The picture that emerges from the current study about radio emission from galaxies is
somewhat different from the one that is obtained by studying the objects contained
in traditional flux-limited radio surveys.  Since the KISS radio sample is a 
quasi-volume-limited sample, it should provide a very good representation of
the overall population of radio-emitting galaxies in the local universe (z $<$ 0.1).
That population is dominated by star-forming galaxies rather than AGNs.  By number,
roughly two thirds of the galaxies in our sample are star-forming galaxies.  In terms of
their integrated radio power, the star-forming galaxies contribute 59\% 
of the total radio emission in the KISS volume.  There are no high luminosity radio
galaxies in this volume, which emphasizes how rare such objects are.  While they
are extremely common in flux-limited radio surveys (e.g., \citet{sad02}, \citet{pran01},
\citet{mag02}) that probe to high redshifts, they make up much less than 1\% of the volume 
density of all radio galaxies.  Another key difference between our sample and these 
radio-selected samples is that the latter typically contain 60\% to 80\% early-type galaxies 
and AGN, and relatively small populations of star-forming galaxies.  

It is possible that our view of the radio galaxy population in the local universe is biased
substantially by the fact that KISS will not be sensitive to elliptical galaxies with weak or
no emission lines.  While we cannot rule this out completely, our belief is that the number 
of radio galaxies within the KISS volume that we are currently missing is actually quite 
small.  First, our RLF agrees well with that of \citet{sad02} in the region where the two
overlap, even though their sample is not biased against early-type radio galaxies.
Second, the Sadler et al. RLF for radio powers above the highest bin covered by
the KISS RLF can be used to predict the number of such galaxies within the KISS
volume (see previous section).  We estimate that no more than 5 -- 6 galaxies are
missing.  Finally, we point out out that at least some elliptical/early-type galaxies {\bf are}
detected by KISS.  For example, KISSR 29 (= NGC 4278) is one such object.  The
presence of many LINERs in the KISS sample also suggests that we have at least
some sensitivity to early-type galaxies with weak emission-line spectra.  Hence, we
feel reasonably confident that, while the KISS radio galaxy sample is probably not
100\% complete, it is probably missing only a modest number of radio-emitting
objects with z $<$ 0.095.

As mentioned previously, we are interested in exploring this issue further.  Therefore,
we have initiated a program of spectroscopy of FIRST radio galaxies with B $<$ 18.5,
in order to fill in the missing radio-emitting galaxies within the KISS volume.  Since there
are no radio-detected KISS ELGs within this volume with B $>$ 18.5, we are confident
that by obtaining these spectra that we will be able to identify all of the remaining
radio galaxies within the KISS volume, and ultimately produce a truly complete,
volume-limited RLF.  If our assessment about missing galaxies above is correct, the
resulting RLF will not look substantially different from the one shown in Figure \ref{fig:RLF}.

The reader will note that we have not discussed two key issues with regard to radio
emission from galaxies.  First, we have not compared our radio fluxes to FIR IRAS
fluxes and constructed a FIR-radio correlation.  Second, we do not attempt to derive
estimates of the star-formation rates of our star-forming radio galaxies.  Both of these
issues will be addressed in a companion paper \citep{cg_iras}, where we consider the
FIR properties of the KISS ELGs and derive star-formation rates using FIR, radio, and
optical (H$\alpha$ flux) methods.

\acknowledgments

We gratefully acknowledge financial support for the KISS project from an NSF 
Presidential Faculty Award to JJS (NSF-AST-9553020), as well as continued 
support for our ongoing follow-up spectroscopy campaign (NSF-AST-0071114).  
We also thank Wesleyan University for providing additional funding for several 
of the observing runs during which the important follow-up spectral data were 
obtained.  We thank the numerous KISS team members who have participated in 
these spectroscopic follow-up observations during the past several years, 
particularly Drew Phillips, Gary Wegner, Jason Melbourne, Laura Chomiuk, 
Kerrie McKinstry, Robin Ciardullo, and Vicki Sarajedini.  Several conversations
with David Helfand during the early stages of this project were extremely helpful.  
We are indebted to the referee for a number of suggestions that improved the final 
presentation of this paper.  We wish to thank the support staffs of Kitt Peak National 
Observatory, Lick Observatory, the Hobby-Eberly Telescope, MDM Observatory,  and Apache 
Point Observatory for their excellent assistance in obtaining both the survey data as 
well as the spectroscopic observations.  Without their assistance this project would 
have been impossible.  Finally, we wish to acknowledge the use of the Digitized Sky 
Survey, which has been produced at the Space Telescope Science Institute under U.S. 
Government grant NAG W-2166.


\clearpage

\renewcommand{\arraystretch}{.6}
\begin{deluxetable}{crrr}
\tablenum{1}
\tablecolumns{4}
\tablecaption{Original and revised FIRST flux values\label{tab:fluxadjust}}
\tablewidth{0pt}
\tablehead{
\colhead{KISS ID} & \colhead{Orig $F_{int}$} & \colhead{New
  $F_{int}$} & \colhead{\% Change}\\
\colhead{} & \colhead{mJy} & \colhead{mJy} & \colhead{}
}
\startdata
    27 &4.19  &  3.94  &  -6.0\\
    71 &26.65 & 27.80  &  +4.3\\
   147 &68.22 & 71.32  &  +4.5\\
   392 &1.43  &  1.73  & +21.0\\
   419 &2.41  &  2.08  & -13.7\\
   439 &2.97  &  2.30  & -22.6\\
   592 &2.96  &  3.94  & +33.1\\
  1125 &6.07  &  6.68  & +10.0\\
  1205 &4.13  &  6.06  & +46.7\\
  1219 &5.96  &  6.25  &  +4.9\\
  1224 &2.03  &  4.96  &+144.3\\
  1561 &3.62  &242.31  &+6594.0\\
  1568 &29.12 & 33.58  & +15.3\\
  1629 &9.73  & 12.54  & +28.9\\
  1673 &4.25  &  3.57  & -16.0\\
  1674 &3.13  &  2.75  & -12.1\\
  1691 &4.98  &  5.31  &  +6.6\\
  1692 &4.78  &  4.57  &  -4.4\\
  1985 &16.51 & 20.18  & +22.2\\
\enddata
\end{deluxetable}


\renewcommand{\arraystretch}{.6}
\begin{deluxetable}{lcccccr}
\tablenum{2}
\tablecolumns{7}
\tablecaption{FIRST and NVSS source matches by ELG type\label{firstnvsstypes}}
\tablewidth{0pt}
\tablehead{
\colhead{} & \multicolumn{5}{c}{ELG Classification} & \colhead{}\\
\colhead{Survey} & \colhead{Sy1(\%)}& \colhead{Sy2(\%)}&
\colhead{SB(\%)}& \colhead{LINER(\%)}& \colhead{QSO(\%)} & \colhead{Total}
}
\startdata
FIRST \& NVSS & 3 (2.4) & 22 (17.7) & 76 (61.3) & 23 (18.5) & 0 (0.0) & 124\\
FIRST only    & 2 (3.3) &  9 (15.0) & 37 (61.7) & 11 (18.3) & 1 (1.7) & 60\\
NVSS only     & 0 (0.0) &  1 (4.3)  & 19 (82.6) &  3 (13.0) & 0 (0.0) & 23\\
Totals        & 5 (2.4) & 32 (15.5) & 132 (63.8) & 37 (17.9) & 1 (0.5) & 207 \\
\tableline\\
High z & 2 & 7 & 0 & 0 & 1 & 10 \\
\enddata
\end{deluxetable}

\clearpage

\renewcommand{\arraystretch}{.6}
\begin{deluxetable}{rcccccrccrcl}
\tablenum{3}
\tablecolumns{12}
\tablewidth{0pt}
\tabletypesize{\tiny}
\tablecaption{Optical and radio properties of the radio ELG sample
  \label{maintable}}
\tablehead{
\colhead {ID}&\colhead {$RA$}&\colhead {$Dec$}&\colhead {$B$}&\colhead
{$M_B$}&\colhead {$L_{H\alpha}$}&\colhead {$S_{1.4}$}&\colhead
{$P_{1.4}$}&\colhead {$z$}&\colhead {$\Delta R$} & \colhead {Survey}&  \colhead
{ELG}\\
\colhead {}&\colhead {hms}&\colhead {dms}&\colhead {}&\colhead
{}&\colhead {erg/s}&\colhead {mJy}&\colhead
{W/Hz}&\colhead {}&\colhead {``}&\colhead {match} &\colhead
{Type}\\
\colhead{(1)}& \colhead{(2)}& \colhead{(3)}& \colhead{(4)}&
\colhead{(5)}& \colhead{(6)}& \colhead{(7)}& \colhead{(8)}&
\colhead{(9)}& \colhead{(10)}& \colhead{(11)} &\colhead{(12)}
}

\startdata
     9 & 12:18:19.3 & 29:15:13.3 & 15.96 & -20.56 &  3.20E41 &      6.41 & 2.88E22 & 0.0479 &     0.61 & B  &Sy2\\
    11 & 12:18:23.4 & 28:58:10.7 & 17.14 & -20.36 &  6.56E41 &      4.95 & 5.46E22 & 0.0746 &     0.81 & B  &SB\\
    27 & 12:19:50.6 & 29:36:52.3 & 11.46 & -19.05 &  1.22E40 &      4.19 & 7.39E19 & 0.0030 &     3.79 & B  &LIN\\
    29 & 12:20:06.8 & 29:16:50.3 & 11.53 & -18.06 &  7.44E39 &    402.00 & 3.08E21 & 0.0020 &     0.40 & B  &LIN\\
    33 & 12:21:34.4 & 28:49:00.4 & 18.24 & -18.81 &  9.83E41 &      9.09 & 6.72E22 & 0.0613 &     0.30 & B  &SB\\
    38 & 12:22:19.5 & 28:49:54.2 & 15.99 & -21.19 &  5.76E41 &      4.67 & 3.89E22 & 0.0649 &     0.20 & B  &LIN\\
    53 & 12:25:28.1 & 29:09:48.5 & 16.60 & -18.93 &  2.06E41 &      0.83 & 1.50E21 & 0.0305 &     1.32 & F  &SB\\
    63 & 12:27:58.8 & 28:49:44.4 & 16.52 & -20.10 &  1.86E41 &      1.55 & 7.58E21 & 0.0500 &     0.83 & B  &LIN\\
    69 & 12:30:26.9 & 28:59:14.3 & 17.25 & -19.73 &  3.63E41 &      0.96 & 6.80E21 & 0.0600 &     0.85 & F  &SB\\
    71 & 12:31:22.9 & 29:08:10.4 & 14.29 & -19.69 &  5.38E41 &     26.65 & 1.19E22 & 0.0152 &     1.32 & B  &LIN\\
    74 & 12:32:43.0 & 29:42:44.4 & 14.68 & -20.56 &  9.73E40 &      3.82 & 5.44E21 & 0.0271 &     9.36 & N  &SB\\
    80 & 12:35:24.0 & 29:29:31.1 & 15.12 & -18.92 &  6.77E40 &      3.09 & 1.42E21 & 0.0155 &    14.00 & N  &SB\\
    84 & 12:37:17.7 & 28:58:39.1 & 16.12 & -20.80 &  1.41E42 &      6.07 & 3.98E22 & 0.0578 &     3.56 & B  &SB\\
    90 & 12:39:14.6 & 29:42:59.0 & 16.80 & -20.13 &  5.84E41 &      1.27 & 8.36E21 & 0.0578 &     0.55 & F  &SB\\
   102 & 12:41:35.1 & 28:50:36.4 & 17.37 & -19.84 &  7.29E40 &     11.59 & 9.94E22 & 0.0659 &     0.86 & B  &LIN\\
   140 & 12:53:05.9 & 29:23:43.3 & 16.66 & -20.37 &  8.95E41 &      1.71 & 1.28E22 & 0.0617 &     0.05 & F  &SB\\
   145 & 12:54:34.8 & 29:36:45.3 & 18.06 & -19.18 &  5.27E41 &      1.45 & 1.32E22 & 0.0678 &     0.69 & F  &SB\\
   147 & 12:54:40.7 & 28:56:17.9 & 12.50 & -20.15 &  3.07E40 &     68.22 & 9.12E21 & 0.0083 &     6.76 & B  &SB\\
   157 & 12:58:09.3 & 28:42:30.6 & 15.53 & -19.54 &  3.12E41 &      5.73 & 7.06E21 & 0.0252 &    10.20 & N  &SB\\
   176 & 13:01:25.1 & 28:40:37.1 & 15.38 & -20.02 &  1.71E41 &      3.91 & 6.47E21 & 0.0292 &    13.40 & N  &SB\\
   177 & 13:01:25.2 & 29:18:48.0 & 15.38 & -19.55 &  7.79E41 &     40.10 & 4.30E22 & 0.0235 &     1.64 & B  &Sy2\\
   179 & 13:01:43.4 & 29:02:40.3 & 14.97 & -19.99 &  1.35E41 &      2.55 & 2.82E21 & 0.0239 &     0.32 & N  &SB\\
   188 & 13:04:22.7 & 28:48:39.0 & 15.42 & -19.77 &  2.11E41 &      3.00 & 4.10E21 & 0.0265 &     9.31 & N  &SB\\
   198 & 13:06:17.3 & 29:03:47.3 & 14.23 & -20.70 &  2.31E41 &     17.62 & 1.91E22 & 0.0237 &     0.98 & B  &LIN\\
   218 & 13:09:16.1 & 29:22:02.8 & 15.38 & -19.29 &  1.60E41 &      4.70 & 4.01E21 & 0.0210 &     0.10 & B  &SB\\
   222 & 13:09:47.5 & 28:54:24.6 & 14.17 & -20.23 &  7.70E40 &      2.72 & 1.80E21 & 0.0185 &     0.31 & B  &SB\\
   227 & 13:11:01.7 & 29:34:41.5 & 15.13 & -19.88 &  3.58E41 &     10.74 & 1.26E22 & 0.0246 &     0.25 & B  &SB\\
   242 & 13:16:03.9 & 29:22:53.7 & 16.71 & -19.23 &  5.29E41 &      1.48 & 4.14E21 & 0.0379 &     1.56 & B  &SB\\
   254 & 13:18:12.3 & 28:45:06.8 & 15.99 & -19.77 &  3.71E41 &      2.36 & 5.60E21 & 0.0349 &     3.01 & N  &SB\\
   266 & 13:19:58.8 & 29:25:38.8 & 15.75 & -19.81 &  4.44E41 &      3.45 & 6.76E21 & 0.0317 &    12.30 & N  &SB\\
   273 & 13:21:56.6 & 28:50:39.7 & 15.54 & -20.28 &  4.01E41 &      3.53 & 8.82E21 & 0.0358 &     1.16 & B  &LIN\\
   285 & 13:25:39.9 & 28:53:39.6 & 17.45 & -19.57 &  2.53E41 &      2.90 & 2.19E22 & 0.0618 &     0.43 & B  &LIN\\
   292 & 13:29:04.1 & 28:58:21.0 & 17.60 & -18.96 &  2.18E41 &      1.58 & 7.82E21 & 0.0502 &     1.24 & F  &SB\\
   332 & 13:40:03.2 & 29:08:14.1 & 16.14 & -20.16 &  2.07E42 &      2.18 & 8.60E21 & 0.0449 &     3.17 & N  &LIN\\
   333 & 13:41:03.0 & 29:36:42.8 & 17.62 & -19.88 &  1.83E42 &      6.44 & 7.70E22 & 0.0776 &     0.47 & B  &SB\\
   337 & 13:41:35.9 & 29:38:11.1 & 17.60 & -19.90 &  1.33E42 &      1.64 & 1.95E22 & 0.0773 &     1.51 & F  &SB\\
   349 & 13:45:52.1 & 29:45:13.3 & 17.45 & -18.62 &  1.82E41 &      3.21 & 1.03E22 & 0.0406 &     0.58 & B  &SB\\
   361 & 13:50:26.7 & 29:25:34.9 & 18.27 & -19.19 &  7.40E41 &      1.02 & 1.18E22 & 0.0762 &     0.58 & F  &SB\\
   363 & 13:50:34.4 & 29:22:22.2 & 16.26 & -19.70 &  1.21E41 &      4.18 & 1.22E22 & 0.0386 &     6.00 & N  &SB\\
   380 & 13:52:19.2 & 29:33:00.8 & 17.44 & -20.03 &  8.48E41 &      1.67 & 1.95E22 & 0.0768 &     0.33 & F  &SB\\
   392 & 13:56:11.3 & 28:59:31.9 & 16.31 & -20.66 &  1.34E41 &      1.73 & 1.26E22 & 0.0609 &     1.36 & B  &LIN\\
   400 & 13:57:21.2 & 28:47:21.8 & 17.15 & -18.72 &  1.36E41 &      6.27 & 1.68E22 & 0.0371 &    21.90 & N  &SB\\
   410 & 13:58:54.9 & 29:34:35.9 & 16.73 & -20.76 &  1.81E42 &      3.11 & 3.69E22 & 0.0773 &     0.25 & B  &SB\\
   419 & 14:00:32.7 & 28:39:38.8 & 16.31 & -20.64 &  7.67E41 &      2.41 & 1.75E22 & 0.0607 &     1.25 & B  &LIN\\
   421 & 14:00:59.2 & 29:33:44.5 & 15.64 & -19.59 &  1.67E41 &      1.92 & 2.83E21 & 0.0276 &     1.11 & B  &SB\\
   422 & 14:01:04.0 & 29:31:31.7 & 15.04 & -20.10 &  3.72E41 &      1.80 & 2.45E21 & 0.0265 &     0.66 & B  &SB\\
   424 & 14:01:29.6 & 29:14:14.1 & 17.85 & -19.23 &  6.17E41 &      0.91 & 7.40E21 & 0.0642 &     2.24 & F  &SB\\
   428 & 14:02:18.7 & 29:44:49.0 & 16.54 & -20.53 &  3.39E42 &      3.22 & 2.59E22 & 0.0639 &     1.77 & B  &SB\\
   429 & 14:02:48.2 & 29:00:44.8 & 17.66 & -19.70 &  1.59E42 &      1.45 & 1.53E22 & 0.0730 &     1.58 & F  &SB\\
   433 & 14:03:43.7 & 29:20:44.5 & 16.46 & -20.60 &  2.42E42 &      1.93 & 1.55E22 & 0.0638 &     0.29 & B  &SB\\
   434 & 14:03:45.1 & 29:21:44.0 & 16.09 & -20.97 &  2.51E42 &      3.90 & 3.11E22 & 0.0636 &     0.50 & B  &Sy2\\
   439 & 14:04:12.9 & 29:39:28.6 & 16.54 & -20.97 &  3.81E42 &      2.97 & 3.59E22 & 0.0780 &     0.64 & B  &SB\\
   466 & 14:09:45.2 & 29:38:38.8 & 18.04 & -19.05 &  1.36E41 &      2.54 & 2.08E22 & 0.0645 &     0.60 & F  &LIN\\
   486 & 14:13:39.6 & 29:34:38.7 & 16.86 & -20.38 &  3.97E41 &      1.63 & 1.53E22 & 0.0689 &     0.36 & F  &SB\\
   497 & 14:15:44.5 & 29:02:20.8 & 17.55 & -20.17 &  7.32E41 &      2.25 & 3.30E22 & 0.0858 &     0.36 & F  &Sy2\\
   501 & 14:16:22.6 & 29:33:04.0 & 16.94 & -20.00 &  1.55E41 &      1.94 & 1.39E22 & 0.0603 &     0.33 & B  &LIN\\
   510 & 14:17:30.3 & 28:47:59.9 & 15.84 & -19.93 &  1.07E42 &      4.44 & 1.08E22 & 0.0353 &     0.46 & B  &SB\\
   511 & 14:17:46.7 & 29:02:41.5 & 16.16 & -21.33 &  7.11E43 &      3.01 & 3.57E22 & 0.0773 &     1.38 & B  &SB\\
   512 & 14:17:48.5 & 29:03:11.2 & 17.37 & -20.14 &  3.73E41 &      3.39 & 4.10E22 & 0.0781 &     0.68 & B  &SB\\
   526 & 14:21:10.0 & 29:10:04.3 & 17.45 & -20.34 &  2.07E42 &      3.26 & 5.12E22 & 0.0887 &     1.03 & B  &SB\\
   531 & 14:22:20.2 & 29:42:55.5 & 15.99 & -20.67 &  8.56E41 &      2.48 & 1.37E22 & 0.0531 &     0.71 & B  &Sy1\\
   546 & 14:25:15.8 & 29:40:24.7 & 17.54 & -19.23 &  2.93E41 &      1.13 & 6.92E21 & 0.0558 &     1.00 & F  &SB\\
   557 & 14:29:04.7 & 29:43:42.6 & 16.97 & -19.73 &  1.80E41 &      2.35 & 1.35E22 & 0.0541 &     0.26 & B  &Sy2\\
   577 & 14:49:07.4 & 29:03:48.0 & 16.97 & -19.71 &  2.60E41 &      1.35 & 7.50E21 & 0.0532 &     1.05 & B  &SB\\
   592 & 14:55:31.1 & 29:27:33.3 & 17.56 & -19.31 &  4.95E41 &      3.94 & 2.50E22 & 0.0568 &     0.18 & B  &SB\\
   612 & 15:01:34.6 & 28:47:14.7 & 17.69 & -19.40 &  5.09E41 &      1.20 & 9.22E21 & 0.0624 &     0.80 & F  &SB\\
   616 & 15:02:22.5 & 29:43:24.1 & 17.81 & -19.57 &  1.50E42 &      0.76 & 7.67E21 & 0.0714 &     0.26 & F  &SB\\
   618 & 15:02:28.8 & 28:58:16.1 & 16.56 & -20.86 &  1.14E43 &      2.88 & 3.00E22 & 0.0725 &     1.38 & B  &Sy2\\
   659 & 15:12:41.2 & 29:15:57.4 & 17.93 & -19.39 &  1.02E42 &      1.93 & 1.90E22 & 0.0705 &     0.80 & F  &SB\\
   707 & 15:22:44.9 & 29:46:10.2 & 15.09 & -19.80 &  4.37E41 &      2.45 & 2.57E21 & 0.0233 &     0.73 & B  &SB\\
   717 & 15:23:48.0 & 28:55:03.6 & 18.04 & -19.76 &  1.11E42 &     12.31 & 1.87E23 & 0.0874 &     0.48 & B  &Sy2\\
   730 & 15:25:17.8 & 29:05:57.8 & 17.94 & -19.63 &  1.28E42 &      1.90 & 2.34E22 & 0.0788 &     0.33 & B  &SB\\
   761 & 15:28:01.6 & 28:59:58.1 & 17.11 & -20.22 &  1.34E42 &      1.95 & 1.93E22 & 0.0708 &     1.08 & B  &SB\\
   762 & 15:28:10.0 & 29:14:36.4 & 18.21 & -19.04 &  6.87E41 &      1.92 & 1.76E22 & 0.0681 &     1.15 & F  &SB\\
   789 & 15:31:31.4 & 29:19:50.7 & 16.90 & -20.39 &  1.19E42 &      1.50 & 1.31E22 & 0.0664 &     0.29 & F  &SB\\
   833 & 15:38:41.2 & 29:27:32.8 & 17.04 & -20.00 &  9.00E41 &      1.43 & 1.04E22 & 0.0609 &     0.33 & B  &SB\\
   834 & 15:38:50.3 & 29:25:26.1 & 16.44 & -20.58 &  1.63E42 &      4.00 & 2.86E22 & 0.0603 &     3.63 & N  &SB\\
   838 & 15:40:11.2 & 29:11:38.7 & 21.28 & -19.71 &  6.67E42 &      1.87 & 5.11E23 & 0.3517 &     0.39 & B  &Sy2\\
   839 & 15:40:22.4 & 29:08:15.0 & 19.74 & -17.99 &  3.02E41 &      0.77 & 1.04E22 & 0.0825 &     0.75 & F  &Sy2\\
   844 & 15:42:27.0 & 29:42:02.6 & 18.43 & -23.17 &  1.38E43 &      1.35 & 6.44E23 & 0.4566 &     0.30 & F  &QSO\\
   872 & 15:50:09.8 & 29:11:07.3 & 16.58 & -21.17 &  1.24E42 &      5.52 & 7.59E22 & 0.0831 &     0.35 & B  &LIN\\
   875 & 15:50:23.0 & 29:01:27.0 & 18.54 & -19.01 &  5.27E41 &      1.61 & 1.84E22 & 0.0758 &     1.51 & F  &SB\\
   896 & 15:54:07.7 & 29:35:29.2 & 17.13 & -20.47 &  2.47E42 &      2.76 & 3.28E22 & 0.0773 &     0.28 & B  &LIN\\
   904 & 15:55:17.9 & 29:06:22.1 & 16.36 & -21.28 &  1.46E42 &      2.07 & 2.53E22 & 0.0785 &     0.50 & B  &SB\\
   916 & 15:56:08.1 & 28:53:10.2 & 17.52 & -19.58 &  4.13E41 &      1.58 & 1.18E22 & 0.0616 &     0.41 & F  &LIN\\
   921 & 15:56:25.7 & 29:04:13.3 & 17.78 & -19.99 &  2.34E41 &      1.53 & 2.09E22 & 0.0829 &     0.37 & F  &LIN\\
   937 & 15:58:48.7 & 28:56:31.9 & 18.50 & -19.15 &  3.01E41 &      2.61 & 3.22E22 & 0.0788 &     1.02 & F  &SB\\
   938 & 15:58:48.8 & 29:54:50.9 & 18.26 & -19.57 &  3.93E41 &      1.48 & 2.15E22 & 0.0855 &     1.19 & F  &SB\\
   946 & 16:00:06.0 & 29:13:33.1 & 17.35 & -20.35 &  9.39E41 &      1.30 & 1.71E22 & 0.0813 &     0.22 & B  &SB\\
   947 & 16:00:19.9 & 29:10:06.9 & 17.02 & -20.59 &  8.63E41 &      2.45 & 2.97E22 & 0.0782 &    17.00 & N  &SB\\
   961 & 16:04:25.4 & 29:24:34.9 & 17.40 & -20.57 &  5.54E41 &      1.35 & 2.35E22 & 0.0934 &     1.04 & B  &SB\\
   967 & 16:06:31.8 & 29:27:57.2 & 17.28 & -20.68 &  6.15E41 &      2.47 & 4.24E22 & 0.0926 &     0.56 & B  &LIN\\
   971 & 16:06:48.1 & 29:10:48.4 & 18.74 & -22.07 &  2.67E42 &      0.74 & 1.76E23 & 0.3290 &     0.98 & B  &Sy1\\
   988 & 16:10:43.8 & 29:18:22.2 & 17.70 & -20.17 &  1.98E43 &      1.37 & 2.12E22 & 0.0882 &     0.48 & F  &SB\\
   997 & 16:12:16.7 & 29:34:23.2 & 17.05 & -19.76 &  5.62E40 &      0.70 & 4.08E21 & 0.0545 &     0.22 & F  &Sy2\\
  1061 & 16:38:34.5 & 29:43:51.0 & 17.88 & -19.96 &  1.11E42 &      1.65 & 2.53E22 & 0.0877 &     0.90 & F  &SB\\
  1078 & 16:47:06.0 & 29:39:18.2 & 17.16 & -19.70 &  6.67E41 &      2.66 & 1.52E22 & 0.0539 &     0.10 & F  &SB\\
  1084 & 16:49:05.3 & 29:45:31.6 & 15.46 & -20.30 &  4.89E41 &      4.96 & 1.03E22 & 0.0327 &     0.47 & B  &SB\\
  1094 & 16:50:47.9 & 28:50:44.5 & 15.44 & -20.38 &  8.19E41 &      3.33 & 7.20E21 & 0.0334 &     1.18 & B  &SB\\
  1097 & 16:52:13.1 & 29:25:01.0 & 17.17 & -20.25 &  3.38E41 &      3.63 & 3.43E22 & 0.0691 &    11.70 & N  &SB\\
  1107 & 16:54:06.8 & 29:21:48.5 & 18.67 & -18.87 &  1.20E42 &      2.66 & 2.79E22 & 0.0727 &     0.50 & B  &SB\\
  1125 & 16:59:08.6 & 28:59:31.1 & 15.10 & -20.79 &  1.12E42 &      6.68 & 1.42E22 & 0.0331 &     0.25 & B  &LIN\\
  1126 & 16:59:20.5 & 29:56:47.1 & 15.02 & -21.28 &  7.57E41 &     16.83 & 5.29E22 & 0.0401 &     0.92 & B  &SB\\
  1136 & 11:54:23.9 & 43:05:09.8 & 15.95 & -21.25 &  2.08E42 &      2.28 & 2.05E22 & 0.0675 &     0.40 & B  &SB\\
  1138 & 11:54:29.4 & 42:58:48.5 & 16.30 & -18.58 &  1.30E41 &      1.12 & 1.19E21 & 0.0235 &     0.61 & F  &Sy2\\
  1140 & 11:54:35.6 & 43:14:57.9 & 16.34 & -20.82 &  7.26E41 &      5.93 & 5.22E22 & 0.0667 &     0.59 & B  &LIN\\
  1146 & 11:55:38.4 & 43:02:44.2 & 14.68 & -20.23 &  2.84E41 &     16.88 & 1.86E22 & 0.0239 &     1.71 & B  &SB\\
  1147 & 11:55:40.6 & 43:48:39.7 & 17.08 & -20.70 &  1.04E41 &      1.22 & 1.89E22 & 0.0881 &     0.97 & F  &Sy2\\
  1150 & 11:55:57.1 & 43:35:34.2 & 15.30 & -20.54 &  1.78E41 &      6.43 & 1.66E22 & 0.0364 &     0.31 & B  &SB\\
  1154 & 11:56:32.9 & 42:59:38.8 & 16.39 & -20.94 &  1.18E42 &      3.09 & 3.43E22 & 0.0720 &     7.77 & N  &Sy2\\
  1155 & 11:57:04.5 & 43:49:37.0 & 16.34 & -18.70 &  1.15E41 &      1.65 & 2.06E21 & 0.0254 &     0.50 & F  &LIN\\
  1165 & 11:58:22.4 & 43:48:54.6 & 16.80 & -20.48 &  6.38E41 &      2.52 & 2.45E22 & 0.0702 &     1.05 & F  &SB\\
  1171 & 11:58:52.6 & 42:34:12.7 & 14.98 & -20.56 &  9.23E41 &     43.01 & 8.45E22 & 0.0318 &     0.76 & B  &Sy2\\
  1205 & 12:04:57.9 & 43:08:58.6 & 15.93 & -20.68 &  6.82E40 &      6.06 & 3.21E22 & 0.0520 &     0.68 & B  &SB\\
  1207 & 12:05:15.3 & 43:15:29.4 & 18.23 & -18.50 &  1.03E41 &      4.89 & 2.90E22 & 0.0549 &     0.75 & B  &Sy2\\
  1219 & 12:09:08.8 & 44:00:11.6 & 14.82 & -21.10 &  1.18E42 &      5.96 & 1.66E22 & 0.0378 &     0.26 & B  &Sy2\\
  1221 & 12:09:17.6 & 44:05:25.0 & 14.92 & -20.97 &  1.14E41 &      4.91 & 1.56E22 & 0.0372 &     4.46 & N  &LIN\\
  1224 & 12:09:26.2 & 44:04:23.9 & 16.05 & -19.88 &  7.40E40 &      4.96 & 1.40E22 & 0.0381 &     0.25 & B  &Sy2\\
  1226 & 12:09:28.0 & 43:16:51.0 & 20.46 & -20.14 &  3.45E43 &     14.54 & 3.04E24 & 0.3096 &     0.19 & B  &Sy1\\
  1241 & 12:13:33.0 & 44:02:34.1 & 17.56 & -19.87 &  1.16E42 &      4.34 & 4.84E22 & 0.0750 &     0.45 & B  &Sy2\\
  1244 & 12:13:55.1 & 42:43:15.6 & 16.48 & -20.18 &  1.32E41 &      1.25 & 6.91E21 & 0.0531 &     0.75 & F  &LIN\\
  1262 & 12:16:55.9 & 44:06:32.1 & 15.84 & -21.30 &  5.04E42 &      5.37 & 4.61E22 & 0.0660 &     0.75 & B  &SB\\
  1274 & 12:18:12.9 & 44:10:22.7 & 14.43 & -20.55 &  2.64E41 &      6.56 & 7.71E21 & 0.0246 &     1.53 & B  &SB\\
  1276 & 12:18:19.9 & 43:53:27.9 & 18.05 & -19.12 &  2.32E42 &      3.13 & 2.77E22 & 0.0669 &     0.94 & B  &SB\\
  1285 & 12:20:53.4 & 42:45:48.8 & 17.84 & -19.58 &  2.58E41 &      1.14 & 1.26E22 & 0.0748 &     1.01 & B  &Sy2\\
  1302 & 12:27:06.5 & 43:30:15.3 & 17.99 & -19.75 &  8.35E41 &      1.91 & 2.85E22 & 0.0865 &     0.64 & B  &SB\\
  1304 & 12:27:41.9 & 44:00:41.5 & 21.61 & -19.26 &  8.37E42 &    382.57 & 1.02E26 & 0.3481 &     0.77 & B  &Sy2\\
  1307 & 12:28:11.9 & 44:05:39.6 & 10.18 & -17.53 &  3.72E38 &    269.00 & 3.61E20 & 0.0006 &     0.50 & B  &SB\\ 
  1313 & 12:29:22.4 & 42:40:15.9 & 18.11 & -18.39 &  2.43E41 &      1.74 & 8.28E21 & 0.0493 &     0.40 & F  &SB\\
  1321 & 12:30:31.6 & 42:58:22.0 & 17.36 & -19.72 &  1.93E42 &      1.92 & 1.56E22 & 0.0642 &     0.56 & F  &Sy2\\
  1342 & 12:34:02.8 & 42:47:17.2 & 17.96 & -19.39 &  4.08E41 &      1.73 & 1.80E22 & 0.0725 &     0.55 & B  &SB\\
  1350 & 12:37:01.4 & 43:38:42.1 & 17.02 & -18.49 &  1.08E41 &      3.00 & 6.07E21 & 0.0312 &     9.32 & N  &SB\\
  1363 & 12:41:48.7 & 43:02:50.3 & 17.37 & -20.02 &  3.70E41 &      1.35 & 1.45E22 & 0.0737 &     0.83 & F  &SB\\
  1382 & 12:49:45.6 & 43:28:56.8 & 16.77 & -20.25 &  8.70E41 &      5.08 & 3.93E22 & 0.0626 &     1.44 & B  &Sy2\\
  1402 & 12:58:58.4 & 43:23:34.4 & 18.16 & -18.77 &  5.47E41 &      1.76 & 1.24E22 & 0.0598 &     1.11 & F  &SB\\
  1406 & 12:59:42.0 & 43:45:11.6 & 15.71 & -21.16 &  1.31E42 &      5.76 & 3.84E22 & 0.0582 &     0.50 & B  &SB\\
  1408 & 13:00:04.2 & 42:42:20.5 & 18.56 & -19.11 &  3.78E41 &      1.37 & 1.91E22 & 0.0838 &     0.23 & F  &SB\\
  1411 & 13:00:25.9 & 43:51:26.3 & 17.50 & -20.00 &  5.85E40 &      1.99 & 2.37E22 & 0.0775 &     1.08 & F  &LIN\\
  1412 & 13:00:48.6 & 43:01:05.2 & 19.09 & -17.83 &  2.62E41 &      7.01 & 4.92E22 & 0.0597 &     0.37 & B  &LIN\\
  1415 & 13:02:04.4 & 42:48:11.9 & 16.80 & -19.89 &  4.01E41 &      2.73 & 1.55E22 & 0.0538 &     0.92 & B  &SB\\
\tableline\\
\\
\\
\\
\\
\\
\\
\\
\\
\\
\\
\\
\\
\\
\\
\\
\\
\\
\\
\\
\\
\\
\\
\\
\\
\\
\\
\\
\\
\\
\\
\\
\\
\\
\\
\\
\\
\\
\\
\\
\\
\\
\\
  1424 & 13:04:08.2 & 44:08:23.8 & 17.40 & -19.48 &  1.39E41 &      7.22 & 4.92E22 & 0.0588 &     0.71 & B  &SB\\
  1439 & 13:05:50.8 & 43:11:38.4 & 16.48 & -18.76 &  7.37E40 &      4.27 & 6.36E21 & 0.0277 &     0.79 & B  &SB\\
  1441 & 13:06:00.4 & 43:46:09.0 & 16.86 & -19.16 &  3.25E41 &     10.56 & 3.24E22 & 0.0396 &     0.75 & B  &LIN\\
  1461 & 13:08:37.9 & 43:44:15.6 & 15.16 & -20.66 &  1.87E41 &     52.27 & 1.33E23 & 0.0361 &     0.37 & B  &LIN\\
  1465 & 13:09:00.7 & 43:34:12.5 & 17.24 & -18.61 &  4.60E40 &      2.30 & 6.04E21 & 0.0367 &     1.09 & F  &LIN\\
  1476 & 13:10:42.4 & 44:01:27.9 & 14.92 & -20.93 &  1.81E40 &      7.62 & 2.00E22 & 0.0367 &     2.99 & B  &LIN\\
  1477 & 13:11:11.2 & 43:43:34.3 & 15.58 & -19.39 &  8.60E40 &      7.61 & 8.83E21 & 0.0245 &     0.91 & B  &SB\\
  1487 & 13:12:59.9 & 43:12:15.5 & 15.57 & -21.44 &  1.36E42 &      9.24 & 6.98E22 & 0.0619 &     0.26 & B  &SB\\
  1494 & 13:13:25.8 & 43:32:14.2 & 15.95 & -20.89 &  1.29E42 &     23.12 & 1.51E23 & 0.0576 &     0.77 & B  &Sy2\\
  1498 & 13:13:54.2 & 44:10:49.6 & 18.15 & -19.30 &  2.83E41 &      2.60 & 2.97E22 & 0.0759 &     0.89 & F  &LIN\\
  1511 & 13:15:10.1 & 43:25:46.8 & 17.58 & -20.16 &  6.94E41 &      1.95 & 2.90E22 & 0.0864 &     0.65 & F  &Sy1\\
  1527 & 13:17:38.0 & 43:48:37.5 & 15.53 & -19.75 &  1.25E41 &      2.66 & 4.13E21 & 0.0283 &     0.82 & B  &SB\\
  1530 & 13:18:33.9 & 43:13:41.0 & 17.39 & -20.08 &  2.43E42 &      4.12 & 4.77E22 & 0.0764 &     1.31 & B  &SB\\
  1533 & 13:18:45.7 & 43:14:06.7 & 16.34 & -20.31 &  1.55E41 &      1.28 & 7.04E21 & 0.0529 &     1.66 & B  &LIN\\
  1537 & 13:19:01.6 & 43:15:00.4 & 16.43 & -20.20 &  1.28E41 &      1.76 & 9.44E21 & 0.0523 &     0.67 & B  &SB\\
  1541 & 13:19:07.3 & 43:08:13.1 & 19.91 & -20.81 &  6.86E42 &      2.23 & 5.19E23 & 0.3259 &     1.15 & F  &Sy2\\
  1554 & 13:22:55.0 & 42:54:36.1 & 17.52 & -19.89 &  7.09E41 &      1.76 & 1.95E22 & 0.0748 &     0.93 & F  &LIN\\
  1556 & 13:23:48.5 & 43:18:03.9 & 15.52 & -19.73 &  4.17E41 &      9.90 & 1.48E22 & 0.0278 &     1.28 & B  &Sy2\\
  1561 & 13:24:04.3 & 43:34:06.4 & 19.43 & -21.39 &  5.00E42 &    242.31 & 6.09E25 & 0.3380 &     0.57 & B  &Sy2\\
  1568 & 13:25:14.0 & 43:16:01.6 & 13.91 & -17.35 &  6.05E40 &     33.58 & 1.27E21 & 0.0044 &     0.54 & B  &SB\\
  1577 & 13:28:43.4 & 43:18:10.9 & 16.28 & -20.98 &  5.68E41 &      3.36 & 3.58E22 & 0.0693 &     7.99 & N  &SB\\
  1578 & 13:28:44.0 & 43:55:50.3 & 15.91 & -19.37 &  2.57E41 &      3.31 & 5.13E21 & 0.0283 &     0.54 & B  &SB\\
  1589 & 13:31:00.4 & 42:50:12.9 & 16.70 & -18.54 &  4.35E40 &      1.03 & 1.54E21 & 0.0277 &     0.84 & F  &SB\\
  1592 & 13:31:56.0 & 44:08:26.1 & 16.22 & -19.46 &  1.30E41 &      6.73 & 1.51E22 & 0.0339 &     1.35 & B  &SB\\
  1600 & 13:34:20.9 & 42:43:08.6 & 21.61 & -19.44 &  4.01E42 &      1.41 & 4.41E23 & 0.3746 &     0.89 & F  &Sy2\\
  1629 & 13:39:35.9 & 43:03:10.0 & 15.22 & -18.21 &  5.03E40 &     12.54 & 3.53E21 & 0.0121 &     1.04 & B  &SB\\
  1631 & 13:39:53.6 & 43:29:01.5 & 16.84 & -20.19 &  5.52E41 &      4.07 & 3.17E22 & 0.0628 &     0.85 & B  &SB\\
  1633 & 13:41:02.0 & 43:57:25.5 & 16.81 & -19.83 &  4.87E41 &      3.11 & 1.69E22 & 0.0526 &     0.80 & B  &SB\\
  1664 & 13:51:29.5 & 43:48:22.2 & 15.38 & -20.29 &  2.75E41 &      7.57 & 1.67E22 & 0.0337 &     1.45 & B  &SB\\
  1666 & 13:52:33.9 & 43:38:33.6 & 16.94 & -19.92 &  6.70E41 &      1.87 & 1.25E22 & 0.0582 &     0.09 & B  &LIN\\
  1673 & 13:54:10.2 & 44:12:48.7 & 16.09 & -20.98 &  6.43E41 &      4.25 & 3.43E22 & 0.0639 &     1.27 & B  &SB\\
  1674 & 13:54:11.9 & 44:12:29.5 & 16.61 & -20.49 &  9.34E42 &      3.13 & 2.60E22 & 0.0648 &     0.99 & B  &SB\\
  1683 & 13:59:02.4 & 42:54:02.2 & 17.48 & -19.88 &  1.01E42 &      1.49 & 1.57E22 & 0.0729 &     0.53 & F  &SB\\
  1691 & 14:00:57.8 & 42:51:20.0 & 16.51 & -19.08 &  4.51E41 &      4.98 & 1.03E22 & 0.0327 &     0.69 & B  &SB\\
  1692 & 14:00:58.8 & 42:50:42.7 & 16.49 & -19.16 &  1.76E42 &      4.78 & 1.04E22 & 0.0335 &     0.34 & B  &SB\\
  1697 & 14:02:26.3 & 43:14:24.6 & 18.26 & -19.50 &  6.62E41 &      2.02 & 3.07E22 & 0.0873 &     1.53 & F  &SB\\
  1743 & 14:13:26.3 & 43:47:08.3 & 17.26 & -20.49 &  6.22E41 &      1.75 & 2.62E22 & 0.0867 &     0.77 & F  &SB\\
  1747 & 14:15:06.4 & 43:48:12.2 & 17.62 & -19.65 &  5.06E41 &      1.81 & 1.74E22 & 0.0697 &     1.12 & F  &SB\\
  1751 & 14:16:47.7 & 42:51:11.6 & 20.15 & -20.72 &  1.29E44 &      6.44 & 1.72E24 & 0.3482 &     0.86 & B  &Sy2\\
  1764 & 14:22:31.0 & 43:42:50.7 & 17.69 & -19.78 &  1.08E41 &      4.51 & 5.27E22 & 0.0767 &     0.57 & B  &SB\\
  1811 & 14:33:47.2 & 43:02:43.4 & 17.65 & -20.15 &  2.05E42 &      1.77 & 2.74E22 & 0.0882 &     0.23 & B  &SB\\
  1835 & 14:39:52.2 & 42:44:32.5 & 14.72 & -18.08 &  1.20E40 &      6.36 & 9.73E20 & 0.0089 &     4.77 & N  &SB\\
  1857 & 14:46:25.0 & 43:49:56.0 & 15.85 & -20.43 &  7.62E41 &      2.16 & 8.29E21 & 0.0443 &     1.58 & B  &SB\\
  1869 & 14:50:39.4 & 42:44:27.2 & 14.53 & -19.86 &  2.04E41 &     10.00 & 6.75E21 & 0.0187 &     0.37 & N  &SB\\
  1871 & 14:51:46.5 & 43:38:40.8 & 14.69 & -18.09 &  3.86E39 &      2.55 & 3.90E20 & 0.0089 &     8.14 & N  &SB\\
  1891 & 15:02:43.3 & 43:16:29.0 & 15.86 & -20.27 &  7.33E40 &      1.62 & 5.37E21 & 0.0412 &     1.00 & F  &LIN\\
  1895 & 15:03:08.7 & 42:38:53.3 & 16.84 & -19.36 &  2.11E41 &      2.33 & 8.21E21 & 0.0425 &     1.34 & F  &SB\\
  1919 & 15:10:09.8 & 43:59:57.6 & 16.93 & -20.27 &  1.38E42 &      2.84 & 2.45E22 & 0.0661 &     0.62 & F  &SB\\
  1930 & 15:15:05.5 & 43:09:02.0 & 15.21 & -19.18 &  3.38E41 &      6.17 & 4.11E21 & 0.0186 &     0.61 & B  &LIN\\
  1961 & 15:32:18.3 & 43:22:30.0 & 17.00 & -19.72 &  1.19E42 &      2.08 & 1.16E22 & 0.0533 &     1.84 & B  &SB\\
  1971 & 15:34:00.8 & 43:40:27.8 & 17.26 & -20.16 &  1.42E42 &      4.92 & 5.23E22 & 0.0732 &     0.13 & B  &SB\\
  1976 & 15:35:04.2 & 43:08:25.4 & 17.34 & -19.73 &  4.16E41 &      2.09 & 1.60E22 & 0.0623 &     1.11 & F  &SB\\
  1980 & 15:36:27.0 & 43:31:07.8 & 14.53 & -20.14 &  4.74E41 &      8.54 & 7.14E21 & 0.0208 &     0.76 & B  &SB\\
  1985 & 15:37:13.2 & 43:17:53.7 & 13.87 & -20.70 &  1.66E41 &     20.18 & 1.54E22 & 0.0199 &     0.95 & B  &SB\\
  1994 & 15:39:07.7 & 43:51:54.9 & 14.82 & -19.72 &  1.59E41 &      5.17 & 3.81E21 & 0.0195 &     0.22 & B  &SB\\
  2009 & 15:43:24.4 & 44:07:17.9 & 15.70 & -20.21 &  3.05E41 &      5.17 & 1.37E22 & 0.0369 &     0.84 & B  &SB\\
  2035 & 15:48:35.3 & 42:41:47.1 & 17.40 & -18.42 &  5.16E42 &      3.82 & 9.50E21 & 0.0357 &     0.96 & B  &Sy2\\
  2073 & 16:01:10.9 & 43:11:38.4 & 18.52 & -18.82 &  1.43E41 &      9.62 & 9.99E22 & 0.0724 &     0.40 & B  &Sy2\\
  2081 & 16:02:16.7 & 42:55:01.4 & 14.96 & -20.05 &  2.14E41 &      3.27 & 3.98E21 & 0.0251 &    10.80 & N  &SB\\
  2097 & 16:05:51.0 & 44:05:40.8 & 21.24 & -19.46 &  1.68E43 &      2.21 & 4.98E23 & 0.3210 &     0.86 & F  &Sy2\\
  2098 & 16:05:58.1 & 44:03:20.1 & 16.20 & -20.09 &  2.81E42 &      2.34 & 9.08E21 & 0.0446 &     1.93 & F  &Sy1\\
  2118 & 16:09:55.5 & 43:07:45.3 & 16.99 & -18.66 &  8.98E40 &      2.27 & 4.97E21 & 0.0328 &     6.23 & N  &LIN\\
  2125 & 16:10:20.4 & 43:00:35.8 & 16.58 & -18.50 &  1.36E41 &      5.64 & 7.29E21 & 0.0258 &     0.67 & B  &SB\\
  2129 & 16:11:08.8 & 44:10:21.8 & 17.67 & -19.27 &  2.84E41 &      2.64 & 1.88E22 & 0.0601 &     0.30 & B  &Sy2\\
  2130 & 16:11:36.4 & 42:45:39.3 & 16.90 & -20.45 &  4.85E42 &      1.28 & 1.33E22 & 0.0725 &     1.02 & B  &SB\\
  2145 & 16:15:31.6 & 42:56:23.3 & 16.53 & -21.07 &  2.10E42 &      1.03 & 1.35E22 & 0.0812 &     0.73 & F  &SB\\
  2148 & 16:15:52.4 & 44:09:40.3 & 16.75 & -20.01 &  2.71E41 &      2.68 & 1.63E22 & 0.0556 &     1.16 & B  &SB \\
\enddata

\end{deluxetable}





\clearpage
\renewcommand{\arraystretch}{.6}
\begin{deluxetable}{lrc|rc|rc}
\tablenum{4}
\tablecolumns{7}
\tablecaption{The local radio luminosity function at
  1.4GHz\label{tab:RLF}}
\tablewidth{0pt}
\tablehead{
\colhead{} & \multicolumn{2}{c}{ALL} & \multicolumn{2}{c}{STARBURST} &
\multicolumn{2}{c}{AGN} \\
\cline{2-7} \\
\colhead{log$_{10}$P$_{1.4GHz}$} & 
\colhead{N} & \colhead{log$\rho_{1.4GHz}$} &
\colhead{N} & \colhead{log$\rho_{1.4GHz}$} &
\colhead{N} & \colhead{log$\rho_{1.4GHz}$}
}
\startdata
20.2 & 1 &  -2.30$^{+0.30}_{-1.0}$  & 0  & \nodata		  & 1 & -2.30$^{+0.30}_{-1.0}$\\ 
20.4 & 0 &  \nodata                 & 0  & \nodata		  & 0 & \nodata \\ 
20.6 & 1 &  -2.20$^{+0.30}_{-1.0}$  & 1  & -2.20$^{+0.30}_{-1.0}$ & 0 & \nodata \\
20.8 & 0 &  \nodata                 & 0  & \nodata                & 0 & \nodata \\
21.0 & 2 &  -2.72$^{+0.23}_{-0.53}$ & 1  & -2.79$^{+0.30}_{-1.0}$ & 1 & -3.52$^{+0.30}_{-1.0}$\\
21.2 & 4 &  -2.78$^{+0.18}_{-0.30}$ & 4  & -2.78$^{+0.18}_{-0.30}$& 0 & \nodata \\
21.4 & 5 &  -3.11$^{+0.16}_{-0.26}$ & 4  & -3.19$^{+0.18}_{-0.30}$& 1 & -3.88$^{+0.23}_{-0.53}$ \\
21.6 & 10 & -3.05$^{+0.12}_{-0.17}$ & 7  & -3.18$^{+0.14}_{-0.21}$& 3 & -3.62$^{+0.18}_{-0.30}$ \\
21.8 & 26 & -2.82$^{+0.07}_{-0.09}$ & 21 & -2.87$^{+0.09}_{-0.11}$& 5 & -3.79$^{+0.16}_{-0.26}$\\
22.0 & 25 & -3.26$^{+0.07}_{-0.09}$ & 17 & -3.48$^{+0.09}_{-0.12}$& 8 & -3.67$^{+0.13}_{-0.19}$\\
22.2 & 47 & -3.10$^{+0.06}_{-0.07}$ & 33 & -3.20$^{+0.07}_{-0.08}$& 14 & -3.79$^{+0.10}_{-0.14}$\\
22.4 & 23 & -3.57$^{+0.08}_{-0.10}$ & 13 & -3.90$^{+0.11}_{-0.14}$& 10 & -3.84$^{+0.12}_{-0.17}$\\
22.6 & 23 & -3.65$^{+0.08}_{-0.10}$ & 14 & -3.85$^{+0.10}_{-0.14}$& 9 & -4.07$^{+0.12}_{-0.18}$\\
22.8 & 10 & -3.97$^{+0.12}_{-0.17}$ & 8  & -4.06$^{+0.13}_{-0.19}$& 2 & -4.72$^{+0.23}_{-0.53}$\\
23.0 & 3 &  -4.55$^{+0.20}_{-0.37}$ & 0  & \nodata                & 3 & -4.55$^{+0.18}_{-0.30}$\\
23.2 & 2 &  -4.74$^{+0.23}_{-0.53}$ & 0  & \nodata                & 2 & -4.74$^{+0.23}_{-0.53}$\\
\enddata
\end{deluxetable}

\renewcommand{\arraystretch}{.6}
\begin{deluxetable}{lcccc}
\tablenum{5}
\tablecolumns{5}
\tablecaption{Schechter Parameters for the KISS and YRC RLFs \label{tab:lf_sol}}
\tablewidth{0pt}
\tablehead{
\colhead{} & \colhead{$L^*$} & \colhead{} & \colhead{$\phi^*$} &\colhead{} \\
\colhead{Fits} & \colhead{($10^{22}$ W/Hz)} & \colhead{$\alpha$} &
\colhead{($10^{-3}$ Mpc$^{-3}$ mag$^{-1}$)} & \colhead{$\chi^2$}
}
\startdata

KISS: All & $4.05\pm0.99$ & $-1.35\pm0.18$ & $1.16\pm0.29$ & 1.66 \\
KISS: SB  & $6.03\pm2.00$ & $-1.52\pm0.16$ & $0.39\pm0.11$ & 3.10 \\
YRC       & $2.64\pm0.12$ & $-1.68\pm0.04$ & $0.49\pm0.03$ & 2.79 \\
\enddata
\end{deluxetable}

\end{document}